\documentclass[a4paper,12pt]{article}
\pdfoutput=1
\usepackage[latin1]{inputenc}
\usepackage[T1]{fontenc}
\usepackage{amsmath}
\usepackage{amsfonts}
\usepackage{amssymb}
\usepackage{color}
\usepackage{graphicx}
\usepackage{caption}
\usepackage{subcaption}
\usepackage{comment}%

\usepackage{color}

 \newcommand{\be}{\begin{equation}}
	 \newcommand{\ee}{\end{equation}}
	 \newcommand{\ba}{\begin{eqnarray}}
		 \newcommand{\ea}{\end{eqnarray}}
		 
		   \newcommand{\bea}{\begin{eqnarray}}
			 \newcommand{\eea}{\end{eqnarray}}
 \newcommand{\nn}{\nonumber}

\begin{document} \title{Non-Abelian vortex lattices}
\author{Gianni Tallarita$^a$ and Adam Peterson$^b$ \\ \vspace{0.2 cm}\\
\normalsize \it $^{a}$ Departamento de Ciencias, Facultad de Artes Liberales, Universidad Adolfo Ibáñez,\\{\normalsize \it Santiago 7941169, Chile.}
 \\
 \\
\normalsize \it $^{b}$ University of Toronto, Department of Physics, Toronto, ON M5S 1A7, Canada.
 \\
}

\date{\hfill}

\maketitle
\begin{abstract}

We perform a numerical study of the phase diagram of the model proposed in \cite{Shifman:2012vv}, which is a simple model containing non-Abelian vortices. As per the case of Abrikosov vortices, we map out a region of parameter space in which the system prefers the formation of vortices in ordered lattice structures. These are generalizations of Abrikosov vortex lattices with extra orientational moduli in the vortex cores. At sufficiently large lattice spacing the low energy theory is described by a sum of $CP(1)$ theories, each located on a vortex site. As the lattice spacing becomes smaller, when the self-interaction of the orientational field becomes relevant, only an overall rotation in internal space survives. 

\end{abstract}
\section{Introduction}

Non-Abelian vortices, first constructed in \cite{Auzzi:2003fs} and \cite{Hanany:2003hp} \cite{Hanany:2004ea}, are Abrikosov-Nielsen-Olesen (ANO) vortices which support additional orientational (non-Abelian) moduli on their world-sheets. They have been widely studied in the literature (see, for example, \cite{Shifman:2003uh} \cite{Gorsky:2004ad} \cite{Eto:2006pg} \cite{Shifman:2004dr} \cite{shifman1} \cite{shifman2} and references therein) as candidates for the vortices responsible for the dual confinement mechanism \cite{Konishi:2002ky} \cite{Konishi:2008vj} \cite{Eto:2006dx}, as originally proposed in \cite{'tHooft:1981ht} \cite{Mandelstam:1974pi}. Although the original models in which they appeared contained a relevant degree of complexity (for example most models involved supersymmetry) recently a particularly simple extension of ANO's original model (based on Witten's superconducting string model \cite{Witten:1984eb}) was proposed which was shown to also contain them \cite{Shifman:2012vv}. In fact, the general idea behind this model was successfully applied to many solitonic solutions with the same outcome: the condensation of a scalar field in the core of the solutions leading to orientational degrees of freedom  \cite{Shifman:2013oia} \cite{Monin:2013kza} \cite{Canfora:2016spb} \cite{Peterson:2015tpa} \cite{Shifman:2015ama} \cite{Peterson:2014nma}\cite{Shifman:2014oqa}. More rencently, the model was also used in a holographic setup to find non-Abelian vortices in a dual 2+1 dimensional superconductor \cite{Tallarita:2015mca}. The appearance of the non-Abelian degrees of freedom has been attributed to distinct sources. In \cite{Forgacs:2016dby}, mainly focusing on cosmic string applications, these were thought to originate from a dark matter sector. In \cite{Peterson:2015tpa}  \cite{Peterson:2014nma} however, a more condensed matter approach was taken and the superconducting system with such additional directional degrees of freedom was shown to have a close analogue in liquid crystals. In particular it was argued that this particular kind of superconductor can be thought of as a superconducting liquid crystal state, with the non-Abelian degrees of freedom playing the part of the director. It remains unclear whether this kind of superconductor can be realised in nature, and if it has any relation to the confining phase of QCD. Therefore, this research can be thought of as a toy model for many physical setups. In a cosmic setup it predicts the existence of periodic arrays of cosmic strings, stabilised by dark matter condensates. From a condensed matter point of view it describes a new phase of matter in which a superconducting liquid crystal form periodic vortex solutions. Finally, from a particle physics perspective it describes the lowest energy configurations of non-Abelian vortices, candidates to be responsible for confinement of quarks.  \newline

The starting setup is the following action, 

\be\label{act1}
S = S_{ANO}+S_{\chi},
\ee
where
\be
S_{ANO} = \int d^4x \left[-\frac{1}{4}F_{\mu\nu}F^{\mu\nu} + (D_\mu \psi)(D^\mu\psi)^*- \lambda\left(|\psi|^2-v^2\right)^2\right],
\ee
\be
S_{\chi}= \int d^4x \left[\partial_\mu\chi^i\partial^\mu\chi^i-\gamma\left(\left(-\mu^2+|\psi|^2\right)\chi^2+\beta \chi^4\right)\right].
\ee
In the above we assume $\lambda, \beta, \gamma >0$ (all of which with mass dimensions zero) and $v >\mu$ (with mass dimension $[+1]$). We work with the conventions
\bea
F_{\mu\nu}&=&\partial_\mu A_\nu - \partial_\nu A_\mu, \\
D_\mu \psi &=& \partial_\mu \psi - i e A_\mu\psi,\\
\chi^2 &=& \chi^i\chi^i,\\
\eta_{\mu\nu} &=& (1,-1,-1,-1). 
\eea

The action enjoys a local $U(1)$ gauge symmetry in the $\psi$ sector and a global $O(3)$ symmetry in the $\chi^i$ sector. We will take $\chi^i$ to be a real triplet field. In recent related work, the additional $\chi$ sector was interpreted as dark matter \cite{Forgacs:2016dby}. The standard quartic potential in the $\psi$ sector breaks the $U(1)$ symmetry spontaneously.  In general there is an interesting range of vacua in this model, the vacuum equations $\partial_\psi V = 0$ and $\partial_\chi V=0$ lead to several branches of solutions
\bea\label{vacua}
\psi = 0, \quad \chi^2 = \frac{\mu^2}{2\beta},\\
|\psi|^2 = v, \quad \chi =0, \\
|\psi|^2 =\frac{\gamma\mu^2-4v^2\beta\lambda}{\gamma-4\beta\lambda},\quad \chi^2 = \frac{2\lambda\left(v^2-\mu^2\right)}{\gamma-4\beta\lambda},
\eea
with the second and the third branch coalescing at the special point $v = \mu$.  We will be interested in the second branch of vacuum solutions which is only a global minimum if 
\be\label{vacreq}
1 < \frac{\gamma}{4\beta\lambda}<(v/\mu)^4. 
\ee
Vacua in which the $\chi$ field condenses are interesting and have been important in specific studies of cholesteric non-Abelian vortices \cite{Peterson:2015tpa} \cite{Peterson:2014nma} but will not be treated here. The condensation of the scalar field ``Higgses'' the photon and gives it a mass 
 \be
 m_A^2 = 2 e^2 v^2,
 \ee 
 defining the penetration depth of the superconductor as $d \approx (1/m_A)$.
 The scalar field is also massive with mass
 \be
 m_\psi^2 = 4\lambda v^2.
 \ee
 This mass defines the coherence length $\xi \approx (1/m_\psi)$ of the superconductor.  The mass of the $\chi$ field is
 \be
 m^2_\chi = \gamma\left(v^2-\mu^2\right).
 \ee
The coupling potential between the two scalar fields, with the assumption that $v>\mu$, leads to $\chi^i$ not condensing in the vacuum.  However, whenever $\psi$ vanishes (in general whenever its value is less than $\mu$), the potential destabilises the $\chi$ field which therefore condenses. In particular this happens in the core of an ANO vortex formed by $\psi$, which is a well known solution of the $\chi^i=0$ theory. This mechanism leads to the existence of non-Abelian vortices, as shown in \cite{Shifman:2012vv}. These vortices are lower in mass than the ANO vortices, so that the condensation of the additional field lowers their energy. The low energy theory describing the orientational gapless excitations is given by a $CP(1)$ non-linear sigma model on the vortex world-sheet. This is most easily seen by the pattern of global symmetry breaking of the $\chi^i$ sector, $SU(2)/U(1)\rightarrow CP(1)$, the remaining $U(1)$ symmetry group corresponding to rotations in the plane defined by the direction of the $\chi^i$ field in internal space (this will be apparent later when we discuss the ansatz).
 
 It is well known however that, for the case of superconductors of type II for which $m_\psi > m_A$, the above system with $\chi^i=0$ has an energy minimizing periodic solution describing a lattice of ANO vortices, each carrying a unit of magnetic flux, the so-called ``Abrikosov lattice''.   \newline
 
This point of transition between superconducting types, and in general whether the condensation of the additional core field, leading to orientational degrees of freedom, supports periodic structures, has not yet been investigated in this model and is essential in order to understand its general behavior over the whole region of parameter space. It is the purpose of this paper to numerically map out the precise nature of the superconducting transition in this model, classifying the superconductor type. The result of the numerical study is Figure \ref{figx},  which shows that in the presence of the $\chi$ field the condition for type II superconductors changes and provides a whole view of the phase transition line plotted against the relevant parameters of the system. In the region marked type II, the system allows periodic vortex structures supporting non-Abelian moduli in their cores, which we also find and present numerically.\newline

  It is important for the rest of the paper to give a short review of Abrikosov's solution for the lattice near the superconducting critical point and of its extension to the full range of magnetic fields (see \cite{Kobayashi:2013axa} for an application to color magnetism). We do this below.\newline
 
The set of coupled equations of motion derived from (\ref{act1}) are (we apply no gauge-fixing condition yet)
\bea\label{eq1}\label{1}
\partial_\mu\partial^\mu\psi -ie(\partial_\mu A^\mu+2 A_\mu\partial^\mu)\psi-e^2 A_\mu A^\mu \psi +2\lambda\psi \left(|\psi|^2-v^2\right)+\gamma\psi \chi^2&=&0,\\
\partial_\mu F^{\mu n}+2e^2 A^n|\psi|^2+ie\left(\partial^n\psi \psi^*-\psi\partial^n\psi^*\right)&=&0,\\
\label{eq3}
\partial_\mu\partial^\mu\chi^n+\gamma\chi^n\left((-\mu^2+|\psi|^2)+2\beta\chi^2\right)&=&0,
\eea
and the energy-momentum tensor functional reads
\bea
T^{\rho\sigma}&=&\frac{1}{4}g^{\rho\sigma}F_{\mu\nu}F^{\mu\nu}-g^{mn}F^{\rho}_m F^{\sigma}_n+2(D^\rho\psi)(D^\sigma\psi)^*-\eta^{\rho\sigma}(D_\mu\psi)(D^\mu\psi)^*\\
&& +2\partial^\rho\chi^i\partial^\sigma\chi^i - \eta^{\rho\sigma}\partial_\mu\chi^i\partial ^\mu\chi^i+\eta^{\rho\sigma}\left(\lambda\left(|\psi|^2-v^2\right)^2+\gamma\left(\left(-\mu^2+|\psi|^2\right)\chi^2+\beta \chi^4\right)\right)\nn.
\eea

We can generally parametrise the fields in order to gauge away the phase of the scalar field. Therefore we pick a parametrization of the form $\psi = f(x,y) e^{i\phi(x,y)}$, where $f(x,y)$ and $\phi(x,y)$ are real,  and $A_\mu= Q_\mu + \frac{1}{e}\partial_\mu \phi$ . From here on we fix the gauge so that $\partial^\mu A_\mu =0$ and consider only static solutions with $A_0= Q_0 =0$. We also define the dimensionless parameters $(\tilde{x}, \tilde{y}) = \frac{m_\psi}{\sqrt{2}}(x,y)$,
\be
a = \frac{m_A^2}{m_\psi^2},\quad b = \frac{\gamma}{4\lambda}\frac{c-1}{c}, \quad c=\frac{v^2}{\mu^2}.
\ee
Note that in terms of these parameters global vacuum stability requires
\be\label{cond2}
\frac{b}{c(c-1)} <\beta,
\ee
for type II superconductors we require $a<1$.\newline
 
We pick an ansatz for the $\chi^i$ field which points in one direction only in the internal space, 
\be
\chi^i = \chi(x,y) (0,0,1).
\ee
Rescaling the dimensionless fields to be $\tilde{f} = f/v$ , $\tilde{\chi} = \sqrt{\frac{2\beta}{\mu^2}}\chi$ and $\tilde{Q} = Q/v$, the dimensionless equations of motion become (we drop all tildes from the fields or derivative operators)
\be
\nabla^2 f- \left[a \left(Q_x^2+Q_y^2\right)+(f^2-1)+\frac{b}{\beta(c-1)} \chi^2\right] f=0,
\ee
\be
\nabla^2 \chi - \frac{2b}{c-1}\left(-1+c f^2+\chi^2\right)\chi=0,
\ee
\be
\nabla^2Q_x -2a\;Q_x f^2 = 0,
\ee
\be
\nabla^2Q_y -2 a\;Q_y f^2 =0.
\ee

For the time being let us ignore the $\chi$ sector (this is equivalent to setting $b=0$). Then, following Abrikosov, we can find periodic solutions to the above equations in the vicinity of $f\approx 0$, which is the point at which superconductivity is destroyed close to the critical magnetic field. In order to do so begin by expanding the fields as the series
\be
f(x,y) = \epsilon f_0 (x,y) + \epsilon^3 f_1(x,y) + \ldots,
\ee
\be
\vec{Q}(x,y) =  \vec{Q}_b (x,y) + \epsilon^2 \vec{Q}_1(x,y) + \ldots 
\ee
where $\epsilon$ is a small parameter denoting the deviation of the applied magnetic field $B$ to the critical one $B_c$ where superconductivity is completely destroyed, $\epsilon = (B_{c}-B)/B_c$. At first order the $\vec{Q}_b$ will give us the applied magnetic field, hence we take this as $ \vec{Q_b} = (-B y,0,0)$, and $B$ is a constant. Abrikosov showed that, in this background and at first order, a periodic solution for the scalar field equation exists for which
\be\label{lattice}
f_0 (x,y) = \bigg|\exp\left(-\frac{e}{2} B y^2\right)\theta_1 \left(\frac{\pi}{x_1}(x+ i y), \frac{x_2 + i y_2}{x_1}\right)\bigg|,
\ee
where $\theta_1$ is the first elliptic theta function and $x_1$, $x_2$ and $y_2$ are parameters which determine the lattice structure. The surface area of the lattice cell is simply $S= x_1 y_2$. For a square lattice $y_2 = x_1$ and $x_2 =0$, for a triangular lattice instead $ y_2 = \frac{\sqrt{3}}{2}x_1$ and $x_2 = \frac{1}{2}x_1$. This result can alternatively be written as a Fourier series as 
\be
\omega_A = f_0^2= \bigg|\sum_{m,n}(-1)^{mn+m+n}\exp\left(\frac{K^2_{mn}}{4e B}\right) e^{i\vec{K}\cdot\vec{r}}\bigg|,
\ee
where 
\be
\vec{K}_{mn}=e B\left(m y_2, -mx_2+nx_1\right), \quad \vec{r}=(x,y).
\ee

The common procedure to extend the lattice solution to the full range of magnetic fields is to use this solution  as a seed to determine the full non-linear structure of the vortex lattice, for an arbitrary magnetic flux. It can be understood as a way to fix the total flux (or equivalently vortex number density) in the action minimization problem, as explained below. Note that for a fixed flux $B$ there are infinitely many solutions corresponding to lattices given by parameters $x_1$ and $y_2$ (nature then presumably chooses these values so as to minimise the energy). \newline

To find the lattice solution for any value of the magnetic flux (and not just those values close to the critical field) the strategy is the following: one must first fix the total flux $\bar{B}$ passing through the integration domain and the lattice structure (given by $x_1$ and $y_2$) and then solves in this background for the local gauge field given by $Q_\mu$ and the scalar field $f$. To do this consider the modified action
\be\label{lag2}
S_{ANO} = \int d^4x \left[-\frac{1}{4}F_{\mu\nu}F^{\mu\nu} -\frac{1}{2} F_{\mu\nu}G^{\mu\nu} + (D_\mu \psi)(D^\mu\psi)^*- \lambda\left(|\psi|^2-v^2\right)^2\right],
\ee
where $G_{\mu\nu}$ is the field strength of an external $U(1)$ gauge field $G_\mu$ whose magnetic flux is 
\be\label{extflux}
\int\int (\nabla\times G)_i dS_i = \bar{B},
\ee
where $S$ here denotes the integration domain (which may comprise more than a single unit cell of the lattice). Therefore $G$ denotes an external constant field, while $F_{\mu\nu}(A)$ denotes the internal electromagnetic response of the superoconductor.

This leads to the equations of motion 

\be\label{othereq}
\nabla^2 f- \left[a \left(Q_x^2+Q_y^2\right)+(f^2-1)\right] f=0,
\ee
\be
\nabla^2Q_x -2a\;Q_x f^2 = \nabla^2 G_x,
\ee
\be
\nabla^2Q_y -2 a\;Q_y f^2 =\nabla^2G_y.
\ee
 It is convenient to use the field variable $Q \rightarrow Q^b_i + G_i$, that is to split the gauge field into the external gauge field plus an internal one \footnote{Note that for the total flux of the cell to be $\bar{B}$, the flux of the field $Q^b_i$ must satisfy
\be
\int\int (\nabla\times Q^b)_i dS_i = 0.
\ee} , so that the equations become 

\be\label{othereq}
\nabla^2 f- \left[a \left((Q^b_x+G_x)^2+(Q^b_y+G_y)^2\right)-(f^2-1)\right] f=0,
\ee
\be
\nabla^2Q^b_x -2a\;(Q^b_x+G_x) f^2 =0,
\ee
\be
\nabla^2Q^b_y -2 a\;(Q^b_y+G_y) f^2 =0.
\ee

Given a fixed flux $\bar{B}$ we must now look for periodic lattice solutions of the above equations with the requirement that $Q^b_i$ has to satisfy a zero flux requirement. A particularly clever way to achieve this is to pick the external gauge field of the form \cite{brandt1} \cite{brandt2} \cite{brandt3}
\be
\vec{G} = -\frac{\nabla \omega_A(\bar{B})\times \hat{z}}{2\kappa\; \omega_A(\bar{B})} + (-\bar{B}y ,0, 0),
\ee
where $\omega_A(\bar{B})$ is given in terms of the original Abrikosov lattice solution for the scalar field (\ref{lattice}) but with $B$ replaced by $\bar{B}$ (this is what we meant when saying that this solution was used as a seed in the full non-linear problem). In this definition $\kappa = d/\xi$ is the ratio of the magnetic penetration depth to the coherence length. For type II superconductors $\kappa \geq 1/\sqrt{2}$. In terms of the units used in this paper we have that
\be
\kappa = \frac{1}{\sqrt{2a}},
\ee
so that the critical $a$ is $a = 1$ corresponding to the point where the scalar and gauge field masses are equal. In terms of these units the critical magnetic fields are $B_{c2} = \kappa$  and
\be
B_{c1} \approx \left(\ln(\kappa)+0.5\right)\frac{1}{2\kappa},
\ee
corresponding to the upper critical magnetic field (field above which superconductivity is completely destroyed) and lower critical fields (approximate field value below which a vortex doesn't want to form) respectively. Note that in these units the critical fields, and as shown later also the flux quantum, depend on $\kappa$ and hence on $a$. This is very important and it amounts to a choice of convention. In order to discuss results at fixed flux quanta we change conventions in a later section. \newline

This definition for $\vec{G}$ allows one to pick the lattice structure by effectively choosing $x_1$ and $y_2$ in the definition of $\omega_A$. The external gauge field $G_i$ is then singular at the vortex positions. This definition satisfies
\be\label{delt}
\nabla \times \vec{G} = \left(\Phi_0  \sum_i^n\delta^{(2)}(r-r_i)\right) \hat{z},
\ee
where $\Phi_0 = \bar{B}/n_v$ ($n_v$ being the number of vortices in the cell) is the single vortex flux  and $\delta^{(2)}$ denotes the two dimensional delta function. The external flux requirement (\ref{extflux}) then trivially follows.  In the above $r_i$ denote the vortex positions in the lattice with coordinates $(x_i, y_i)$.  Note that this definition does not restrict $\bar{B}$ to be in the vicinity of the critical flux $B_c$. It rather defines a vortex lattice structure with the appropriate magnetic field singularities located at the vortex cores, supporting a total flux $\bar{B}$. The delta function originates from the fact that close to the vortex cores $\omega_A \approx \tilde{r}^2$, where $\tilde{r}=r-r_0$ and $r_0$ is the coordinate position of the vortex core, so that $\vec{G} \approx\frac{\vec{\hat{z}}\times\vec{\tilde{r}}}{\tilde{r}^2}$ and we have the important result that 
\be\label{delta}
\nabla\times \frac{\vec{\hat{z}}\times\vec{\tilde{r}}}{\tilde{r}^2} = 2\pi\delta^{(2)}(\tilde{r}).
\ee

Therefore, for a particularly chosen lattice at fixed $\bar{B}$, once $x_1$ is specified and its relation to $x_2$ and $y_2$ is fixed (and therefore $\omega_A(\bar{B})$ is known), this can be used to determine the local gauge field $Q^b_i$ appearing in equations (\ref{othereq}) from a numerical minimization problem. In terms of the parameters of our model, the external total flux over the integration domain is given by $\bar{B}= 2\pi n/\kappa$, therefore the magnetic field carried by each unit cell is $B= \frac{2\pi}{x_1\;y_2\; \kappa}$ where $n$ counts the number of lattice cells in our numerical integration domain. \newline 

\begin{figure}[ptb]
\begin{subfigure}{.5\textwidth}
\centering
\includegraphics[width=0.8\linewidth]{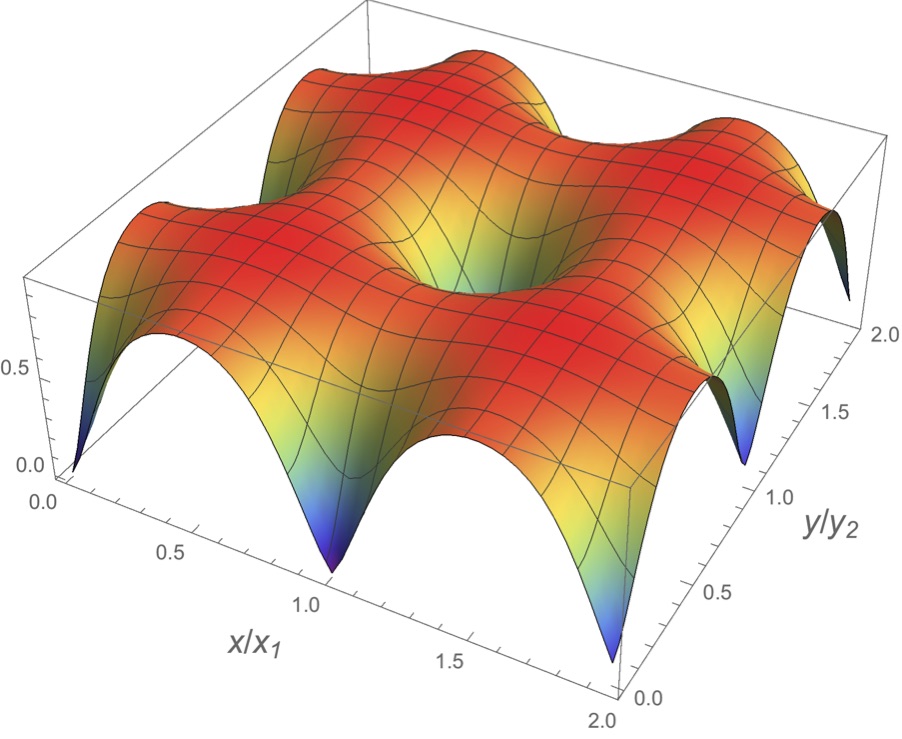}
\caption{$f$}
\end{subfigure}
\begin{subfigure}{.5\textwidth}
\centering
\includegraphics[width=0.8\linewidth]{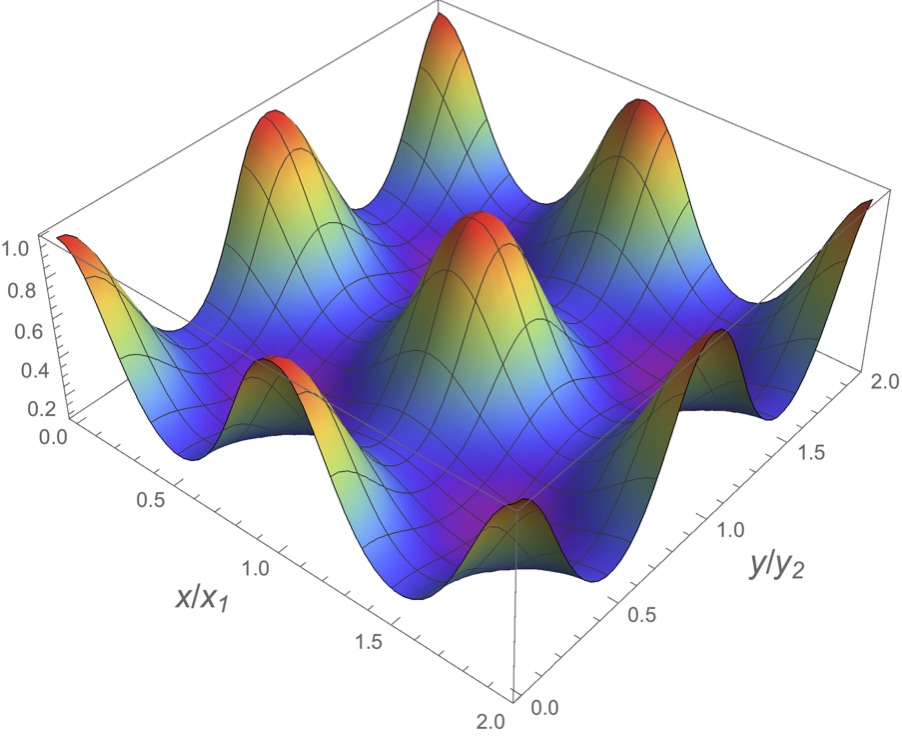}
\caption{$B$}
\end{subfigure}
\begin{subfigure}{.5\textwidth}
\centering
\includegraphics[width=0.8\linewidth]{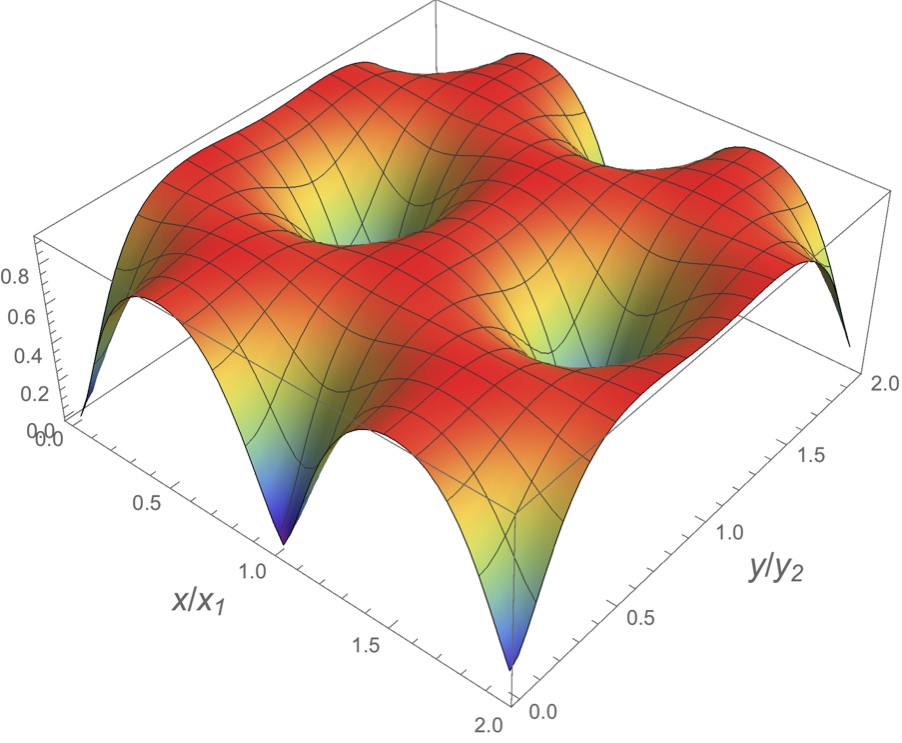}
\caption{$f$}
\end{subfigure}
\begin{subfigure}{.5\textwidth}
\centering
\includegraphics[width=0.8\linewidth]{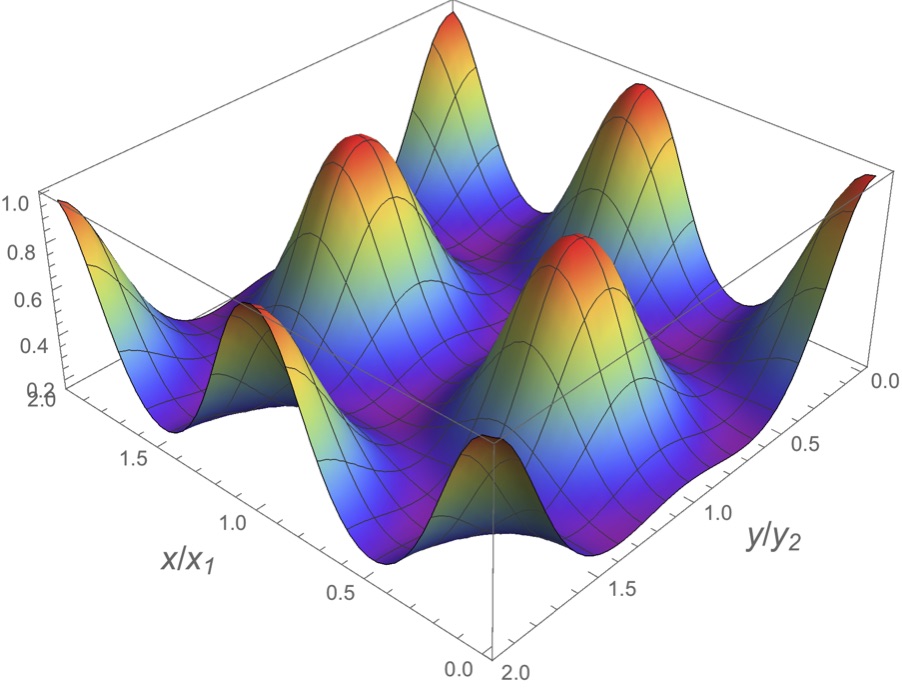}
\caption{$B$}
\end{subfigure}
\caption{Vortex Lattice plots with $\chi=0$ for a square lattice (top row) and triangular lattice (bottom row) with $x_1 = 5$ at $a =0.9$.}%
\label{fig1}%
\end{figure}

Some model solutions are shown in figure \ref{fig1}. In these figures $B$ is the magnetic field of $Q_i$ whose total flux is $\bar{B}$. The solutions represent Abrikosov Lattices for a generic flux $\bar{B}$ per unit cell of both square and triangular geometries. The energy of these solutions is discussed in a later section.

\subsection{Numerical procedure}

The equations are solved numerically using a relaxation procedure. The derivatives are discretized using a second order central finite difference method. The accuracy of the procedure is $\mathcal{O}(10^{-6})$. The results are further checked using a Newton-Rhapson method, which reproduced the same results. We impose periodic boundary conditions at the borders of the lattice and demand that the scalar field $f$ vanish at the vortex centers.

\section{Non-Abelian Vortex Lattices}\label{int}

It is the purpose of this paper to investigate the extension of the above setup to include the condensation of the additional $\chi$ field. At large enough lattice spacing the condensation of the $\chi$ field should resemble that of isolated non-Abelian vortices. However, as the lattice spacing is reduced, these self interactions will become relevant and the structure of the lattice must change. 

\subsection{Non-Abelian vortex - vortex forces}\label{int}

Let us discuss briefly the relevant aspects of the non-Abelian vortex interactions which are important for the rest of the paper. \newline

The vortex-vortex interaction can most easily be calculated by an analysis of the far-distance behaviour of the fields and an assumption of large vortex separation (we will follow the discussion in \cite{Manton:2004tk}). We restrict our considerations to the case of largely separated non-Abelian vortex orientations (a general case can be found in \cite{Auzzi:2007wj}).  If we consider the case of an isolated non-Abelian vortex, we must go back to equations (\ref{eq1}) - (\ref{eq3}) and, switching to cylindrical coordinates with radial variable $\rho$ use the ansatz

\be
A_i = -\epsilon_{ij}\frac{x_j}{\rho^2}(n_e- Q(r)),
\ee
\be
\psi = f e^{i n_e \theta},
\ee
\be
\chi^i = \chi(\rho)^i,
\ee
all of which are understood in a-dimensional units. Here $n_e$ is the quantum of flux carried by each vortex. With these ansatz the field equations reduce to
\be\label{chieqs1}
\partial_r^2f+\frac{1}{r}\partial_rf- \left[\frac{Q^2}{r^2}+(f^2-1)+\frac{b}{\beta(c-1)} \chi^2\right] f=0,
\ee
\be
\partial_r^2\chi^i+\frac{1}{r}\partial_r\chi^i - \frac{2b}{c-1}\left(-1+c f^2+\chi^2\right)\chi^i=0,
\ee
\be\label{chieqs3}
\partial_r^2Q-\frac{1}{r}\partial_rQ -2a\;Q f^2 = 0.
\ee

Note that we have rescaled the field $Q \rightarrow Q/\sqrt{a}$ compared to (\ref{othereq}), this rescaling is important in discussing the phase diagram in a later section \footnote{This is simply the change of convention where the gauge coupling $e$ appears in the denominator of the field strength term in the action, rather than in the covariant derivative}.  This change of convention is important as it removes the dependence on $a$ in the flux quanta. Therefore it allows us to change the parameter $a$ appearing in the equations of motion without changing the flux of the vortices. Throughout the paper we will use the first convention when displaying solutions to the lattice system (this is done in order to remain in contact with \cite{brandt1}) and the latter convention when discussing single vortex results (such convention is adopted in \cite{Shifman:2012vv}, for example). Either convention is especially simple in discussing both aspects separately, and hence we choose to switch between both when needed.\newline

The magnetic field is $B = \frac{1}{r} Q'$, with $'$ denoting differentiation w.r.t $r$. Linearising the fields around their leading behaviours far from the vortex cores,
\be
f \approx 1-\sigma(\rho),\quad Q\approx 0, \quad \chi^i\approx 0
\ee
the resulting leading behaviours are given by the usual modified Bessel functions

\be
\sigma(\rho)\approx c_1 K_1(\sqrt{2}\rho),\quad Q\approx c_2 \rho K_0(\sqrt{2a}\rho), \quad \chi^i\approx c_3^i K_1(\sqrt{2b}\rho)
\ee

\noindent where $c_i$ are integration constants, which have leading order expansions at large $\rho$ of the form

\be\label{expansions}
\sigma(\rho)\approx \frac{\tilde{c}_1}{\sqrt{\rho}}\; e^{(-\sqrt{2}\rho)},\quad Q\approx \tilde{c}_2 \sqrt{\rho}\; e^{-(\sqrt{2a}\rho)}, \quad \chi^i\approx \frac{\tilde{c}_3^i}{\sqrt{\rho}}\; e^{-(\sqrt{2b}\rho)}
\ee

\noindent where $\tilde{c}_i$ are in general different constants from $c_i$ (these are obtained from numerical integration of the vortex profiles). As expected, the scalar field $\chi$ has a general behaviour which is very similar to the scalar field $f$ in terms of its exponential decay. 

The magnetic field is $B = \frac{1}{r} Q'$, with $'$ denoting differentiation w.r.t $r$. The corresponding expression for the energy of the isolated vortex is 
\be\label{totE}
E = E_f + E_\chi,
\ee
where
\be
E_f = 2\pi\int dr\; r \left[(\partial_r f)^2 +\frac{1}{2r^2}(Q')^2+\frac{1}{2}\left(f^2-1\right)^2+\frac{a}{r^2} f^2Q^2\right],
\ee
and
\be
E_\chi = 2\pi\int dr \;r \left[\frac{1}{2c\beta}(\partial_r \chi)^2+\frac{1}{\beta}\frac{b}{c(c-1)}\left(\frac{1}{2}\chi^4-\chi^2(1-c f^2)\right)\right].
\ee

The vortex-vortex interaction energy $E_{int}$ at large vortex separations,  is generally calculated  by subtracting from the total energy of a vortex-vortex configuration the energy of the two isolated vortices. To do so one must first take a vortex-vortex solution ansatz. The good ansatz is given by 
\be
f = f_1\times f_2, \quad Q = Q_1 + Q_2, \quad \chi^i = \chi_1^i + \chi_2^i,
\ee
where the subscript on the fields indicates they are the field profiles of vortex 1 and 2 respectively. Therefore, to conserve the right topological properties one must take the product of the scalar field profile $f$. Then, at large vortex separations, it is sufficient to consider only the far field behaviours of the fields in the energy,
\be
f \approx (1-\sigma_1-\sigma_2), \quad Q \approx Q_1 + Q_2, \quad \chi^i \approx \chi_1^i + \chi_2^i,
\ee
where we understood the field profiles to be those given by equations (\ref{expansions}) in the above. Then, inserting these expressions into (\ref{totE}) and subtracting the energies of each isolated vortex we find, at leading order,

\be
E_{int} = 2\pi \int d\rho \rho \left(\frac{2}{\rho^2}Q_1' Q_2' + 2\sigma_1'\sigma_2' +\frac{1}{c\beta} \chi'_1\cdot\chi'_2+\frac{2a}{\rho^2}Q_1 Q_2 + 4\sigma_1\sigma_2 +\frac{2b}{c\beta} \chi_1\cdot\chi_2\right).
\ee

Using the field expansions (\ref{expansions}), assuming similar profiles for all the $\chi^i$, and the techniques outlined in \cite{Bettencourt:1994kf} this expression can be integrated to give
 \be
 E_ {int} = - \tilde{c_1}^2 K_0(\sqrt{2}s) - \hat{c}\cdot \hat{c} K_0(\sqrt{2b}s) + \tilde{c_2}^2 K_0(\sqrt{2a}s),
 \ee
 where, $\hat{c}$ is the constant vector of constants appearing from the integration of $\chi^i$ and $s$ is the vortex separation assumed to be large. If $\chi=0$, in the standard Abrikosov vortex case, when $a=1$, at the so called critical point, $\tilde{c_1} = \tilde{c_2}$ and therefore $E_{int}=0$. In the presence of the $\chi$ field, which is after all adding a scalar sector, we have an additional force channel between the vortices. When the vortices have parallel internal orientations, then $\hat{c}\cdot \hat{c} > 0$ and the force is attractive. This attractive mechanism was also observed separately in a numerical study of Skyrmions \cite{Canfora:2016spb}. When the vortices are antiparallel, $\hat{c}\cdot \hat{c} < 0$ and the force is repulsive. When they are completely transverse instead, the $\chi$ field interaction vanishes. In particular, it is no longer true that at $a=1$ critical vortices have zero interaction energy (we expect that at this order however $\tilde{c_1} = \tilde{c_2}$ should remain true). This means that the condition on type II vortices and the point of criticality is generally expected to be different than before. In field terms, there is no BPS condition at $a=1$ even though it doesn't exclude the possibility that there is one since $E_{int}$ might still vanish in general for specifically chosen values of the parameters. This should now be determined numerically from the full field profiles. We will not perform a numerical investigation on the values of the parameters $\tilde{c}_i$ in this paper as we generally solve the full field equations. However, this would be an interesting avenue of research per se. \newline
 
Note also that the change in nature of the interaction force between non- Abelian vortices with respect to their internal orientation is indicative of the possibility of alternative (as in besides the parallel orientation case) lattice structures. We will devote a section of this paper to the exploration of such structures.

\subsection{Solutions}

Now we wish to find the full 2D solutions which describe lattices of non-Abelian vortices. We therefore switch back to the conventions of (\ref{othereq}). In the presence of the $\chi$ field the equations we must solve are

\be\label{othereq2}
\nabla^2 f- \left[a \left((Q^b_x+G_x)^2+(Q^b_y+G_y)^2\right)+(f^2-1)+\frac{b}{\beta(c-1)} \chi^2\right] f=0,
\ee
\be
\nabla^2 \chi - \frac{2b}{c-1}\left(-1+c f^2+\chi^2\right)\chi=0,
\ee
\be
\nabla^2Q^b_x -2a\;(Q^b_x+G_x) f^2 =0,
\ee
\be
\nabla^2Q^b_y -2 a\;(Q^b_y+G_y) f^2 =0.
\ee

\begin{figure}[ptb]
\begin{subfigure}{.5\textwidth}
\centering
\includegraphics[width=0.8\linewidth]{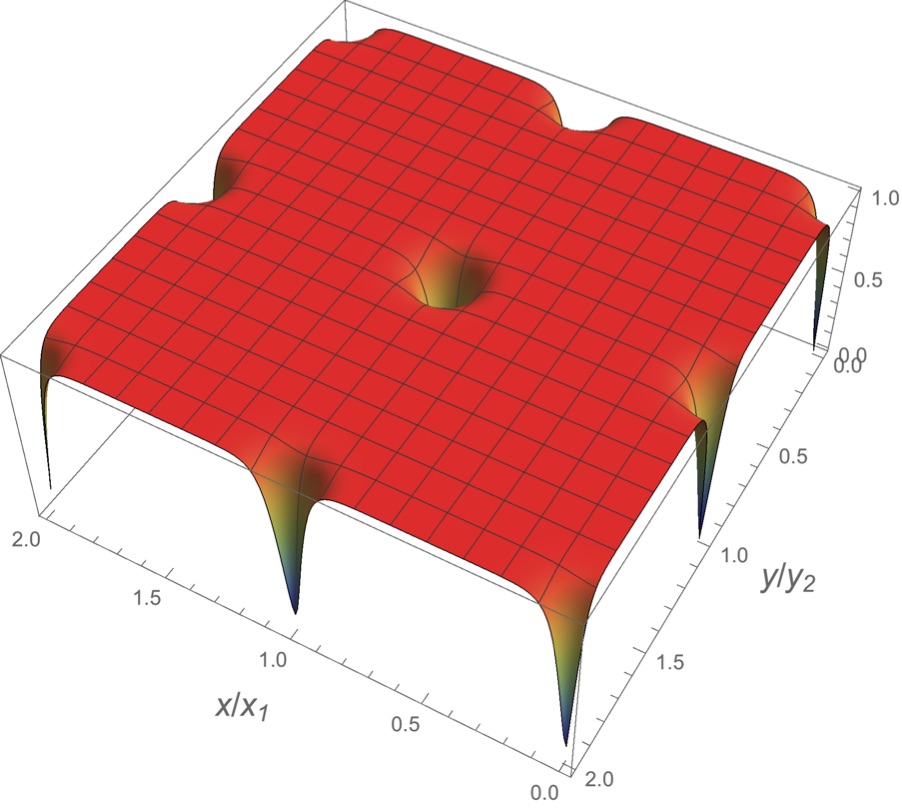}
\caption{$f$}
\end{subfigure}
\begin{subfigure}{.5\textwidth}
\centering
\includegraphics[width=0.8\linewidth]{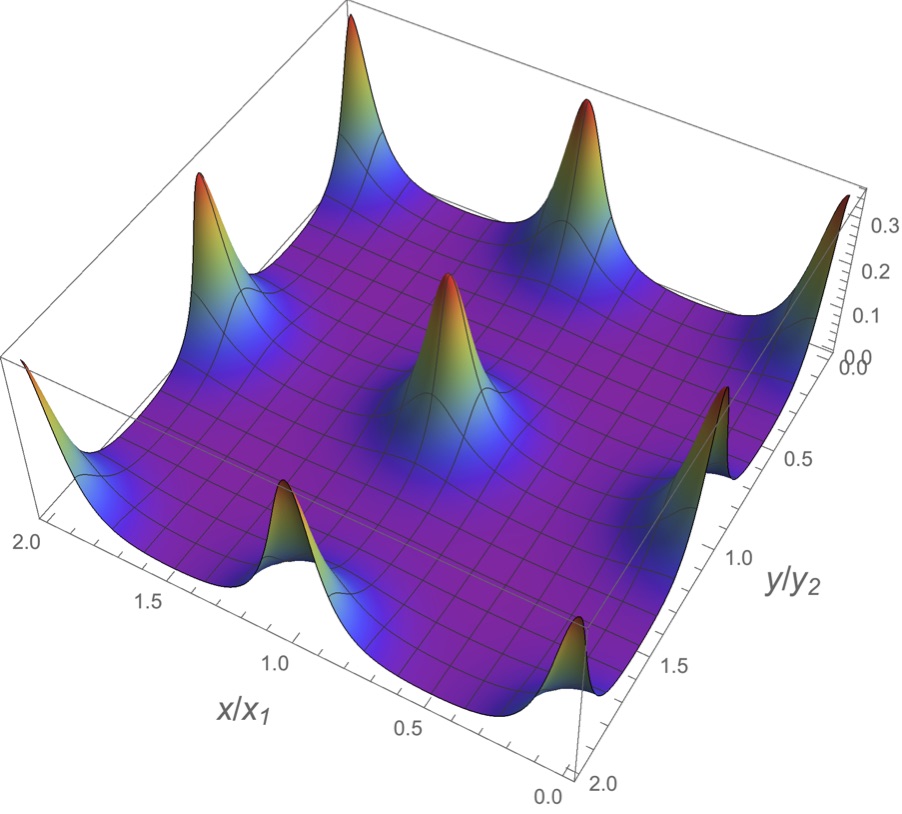}
\caption{$\chi$}
\end{subfigure}
\center
\begin{subfigure}{.5\textwidth}
\centering
\includegraphics[width=0.8\linewidth]{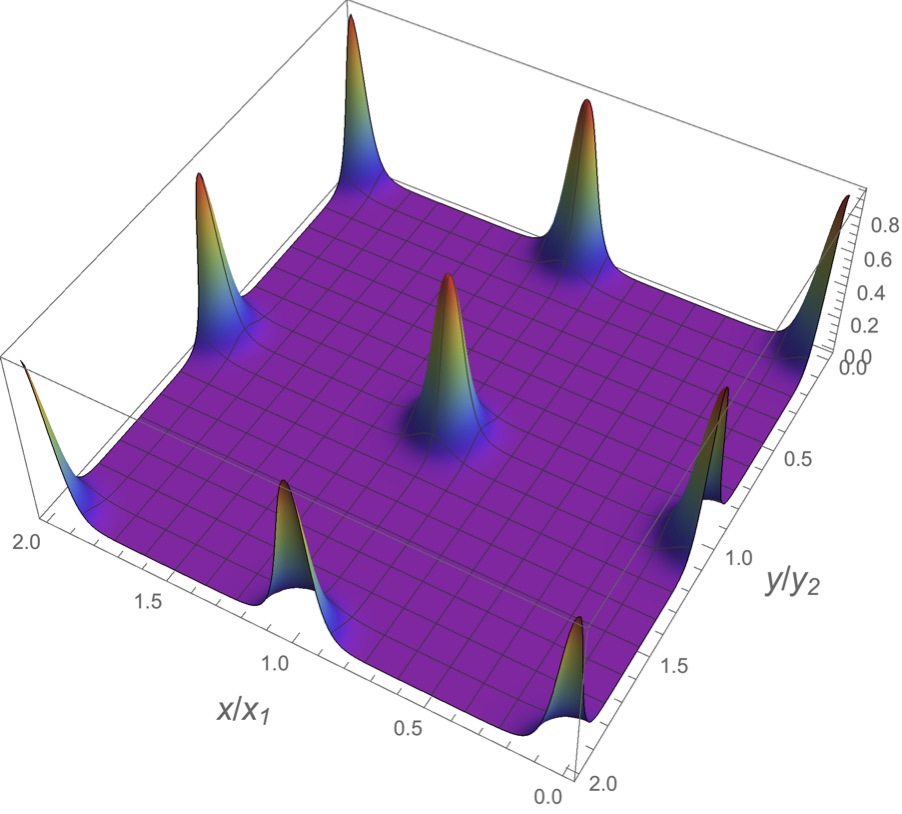}
\caption{$B$}
\end{subfigure}
\caption{Non-Abelian Square Vortex Lattice  $x_1 = 20$ at $a =0.9$, $b=0.1$, $c=1.2$ and $\beta = 1.4 b/(c(c-1))$.}%
\label{fig2}%
\end{figure}

\begin{figure}[ptb]
\begin{subfigure}{.5\textwidth}
\centering
\includegraphics[width=0.8\linewidth]{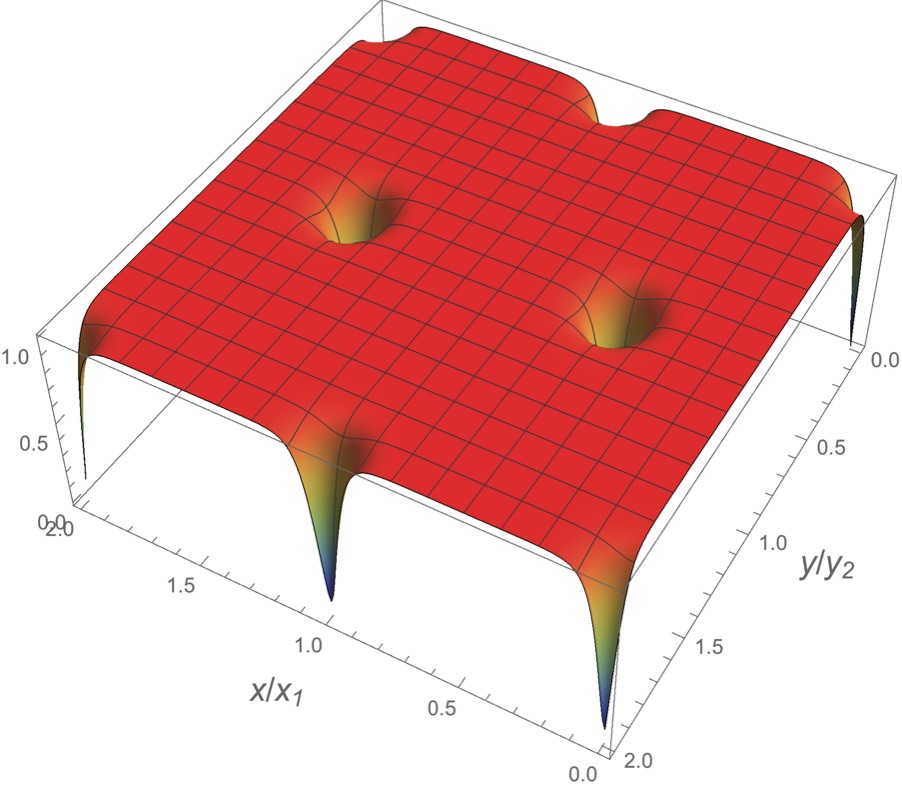}
\caption{$f$}
\end{subfigure}
\begin{subfigure}{.5\textwidth}
\centering
\includegraphics[width=0.8\linewidth]{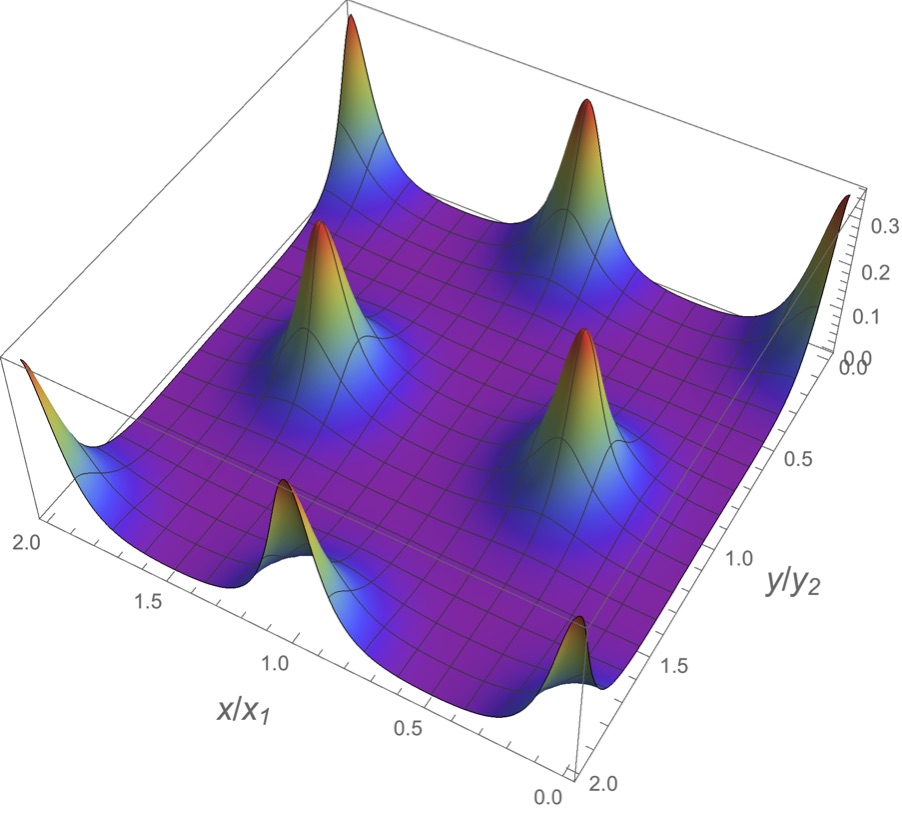}
\caption{$\chi$}
\end{subfigure}
\center
\begin{subfigure}{.5\textwidth}
\centering
\includegraphics[width=0.8\linewidth]{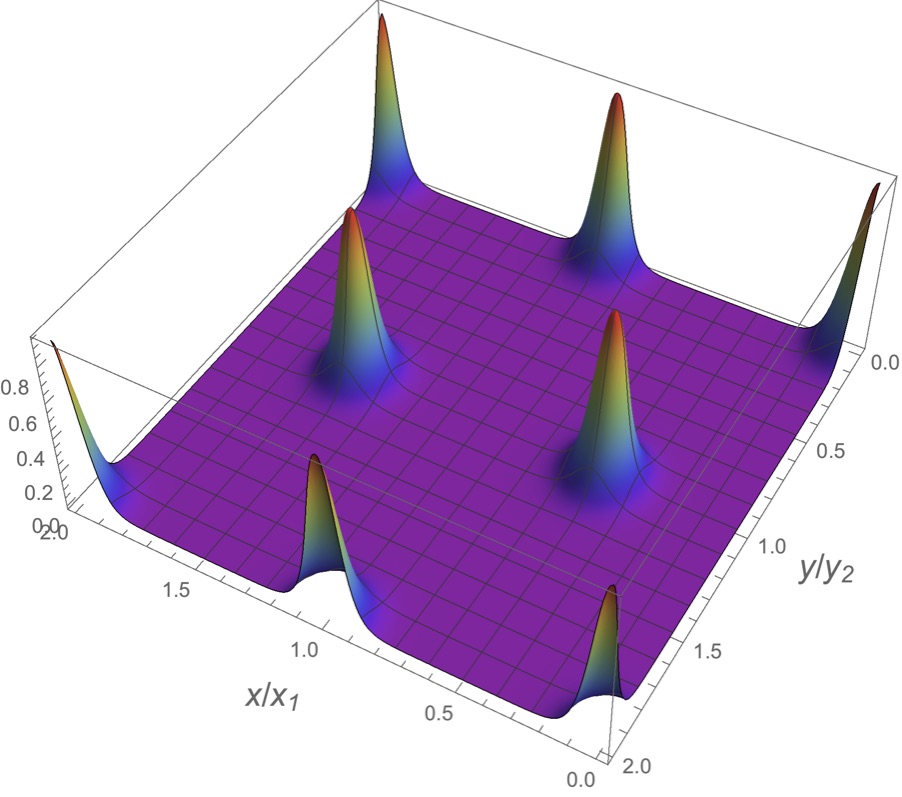}
\caption{$B$}
\end{subfigure}
\caption{Non-Abelian Triangular Vortex Lattice  $x_1 = 20$ at $a =0.9$, $b=0.1$, $c=1.2$ and $\beta = 1.4 b/(c(c-1))$.}%
\label{fig3}%
\end{figure}

\begin{figure}[ptb]
\begin{subfigure}{.5\textwidth}
\centering
\includegraphics[width=0.8\linewidth]{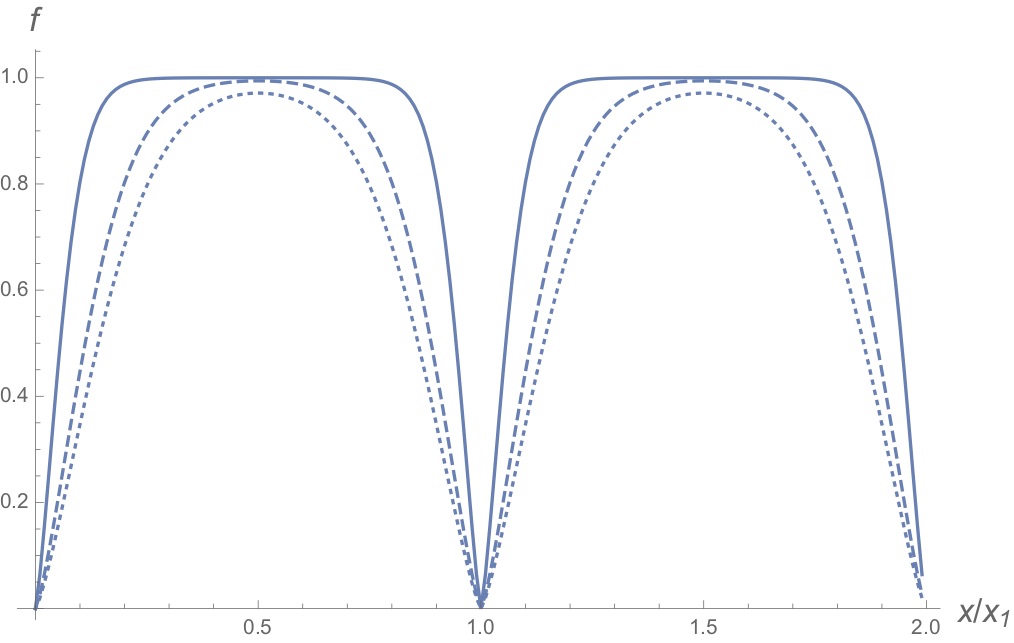}
\caption{}
\end{subfigure}
\begin{subfigure}{.5\textwidth}
\centering
\includegraphics[width=0.8\linewidth]{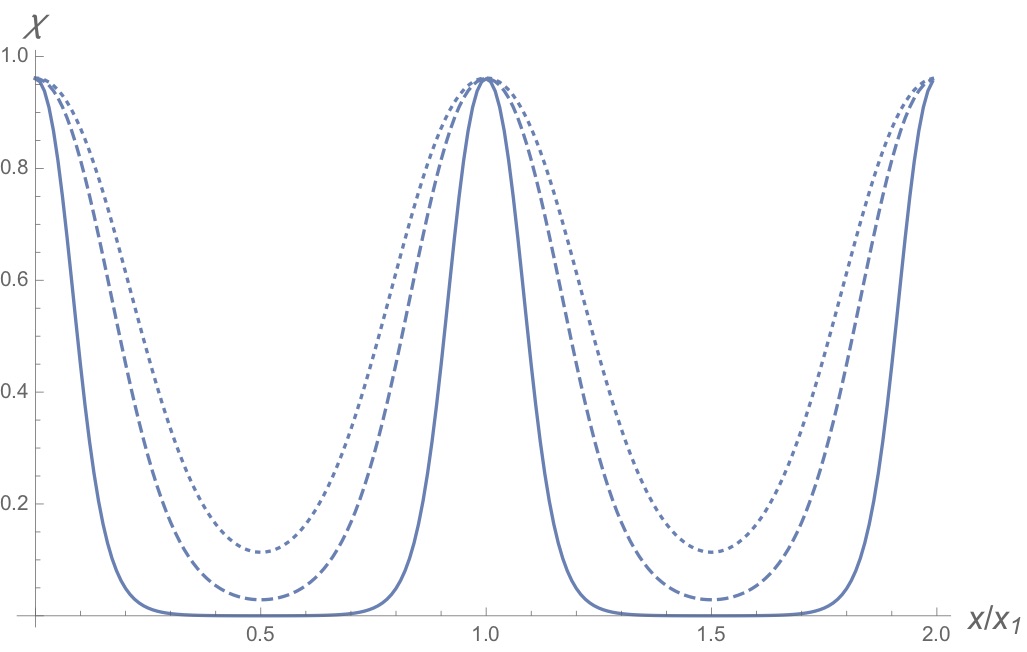}
\caption{}
\end{subfigure}
\caption{Field Profiles for $f$ and $\chi$ are shown at $y=0$ for the square lattice with various vortex spacings.  The profiles for the spacings $x_1 = 20$, $10$, and $8$ are shown in solid, dashed, and dotted curves respectively.}%
\label{fig4}%
\end{figure}

\begin{figure}[ptb]
\begin{subfigure}{.5\textwidth}
\centering
\includegraphics[width=0.8\linewidth]{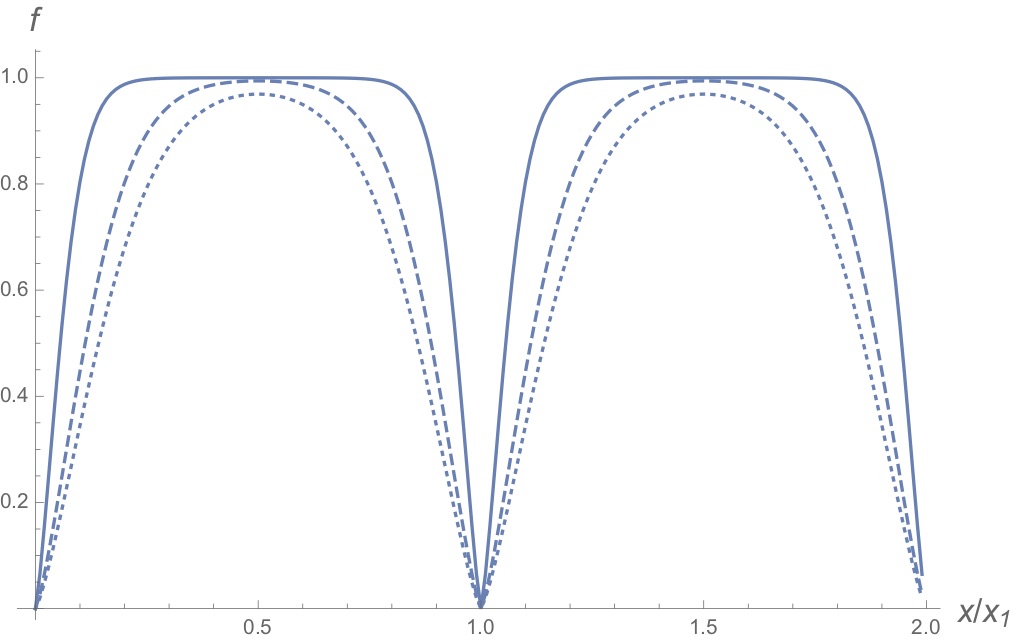}
\caption{}
\end{subfigure}
\begin{subfigure}{.5\textwidth}
\centering
\includegraphics[width=0.8\linewidth]{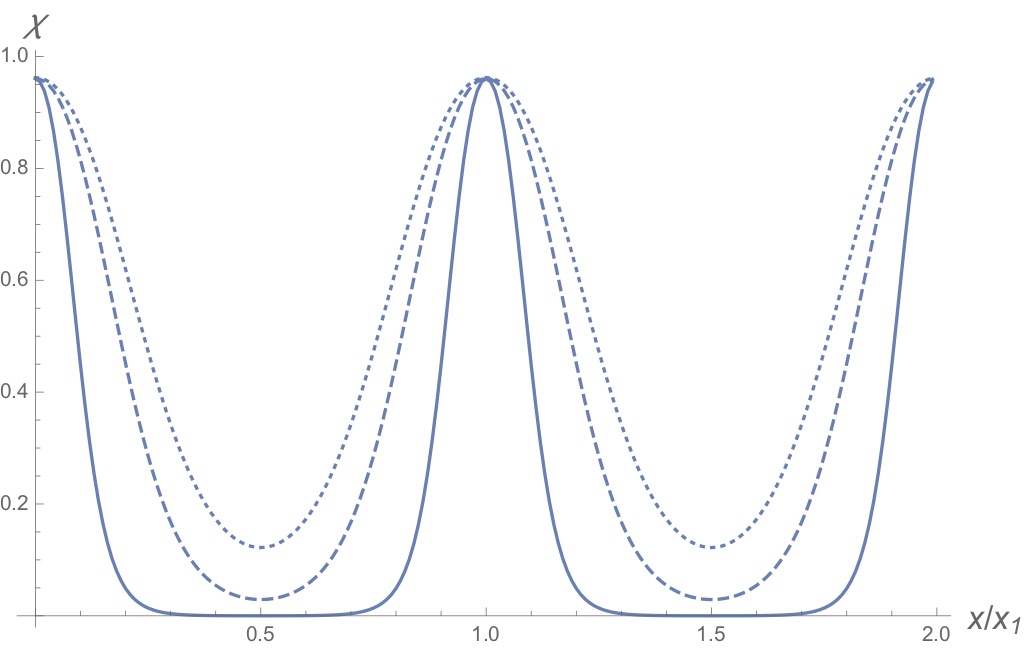}
\caption{}
\end{subfigure}
\caption{Field Profiles for $f$ and $\chi$ are shown at $y=0$ for the triangle lattice with various vortex spacings.  The profiles for the spacings $x_1 = 20$, $10$, and $8$ are shown in solid, dashed, and dotted curves respectively.}%
\label{figtr}%
\end{figure}

We use an identical numerical procedure as outline above in order  to find the solutions. Figures \ref{fig2} and \ref{fig3} show solutions at large lattice spacing (of square and triangular geometries respectively). These represent a lattice of isolated non-Abelian vortices, with the $\chi$ field condensing only in the core of the Abrikosov vortices and quickly decaying to zero outside. As we bring the vortices closer together by decreasing the lattice spacing, self-interactions of the $\chi$ field become important. The field gradually and smoothly lifts and becomes non-vanishing over the whole lattice (see figure \ref{fig4} and figure \ref{figtr}). These solutions do not represent a lattice of non-Abelian vortices, in this case the $\chi$ field is non zero over the whole lattice, which implies the internal orientational moduli are delocalised from the vortex cores.

\begin{figure}[ptb]
\centering
\begin{tabular}{|c|c|c|c|}
\hline
$a$&$x_1$&$E_{\square}$&$E_{\triangle}$\\ \hline
0.9&5&6.78&6.57\\\hline
0.9&2.5&3.13&2.74\\\hline
0.7&10&6.38&6.34\\\hline
0.4&5&3.51&3.44\\\hline
0.4&2.5&2.56&2.39\\\hline
\end{tabular}
\caption{Representative energies at different parameters and lattice spacings for square or triangular lattices of the Abrikosov type.}%
\label{figtable}%
\end{figure}

\subsection{Anti-parallel configurations on the square lattice}

\begin{figure}[ptb]
\begin{subfigure}{.5\textwidth}
\centering
\includegraphics[width=0.8\linewidth]{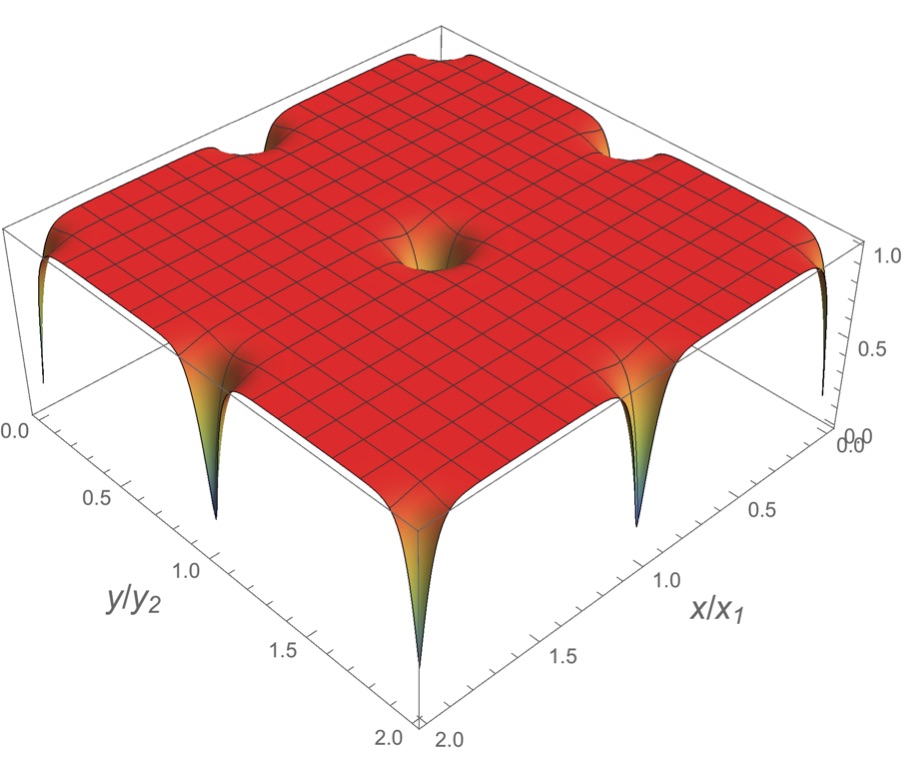}
\caption{$f$}
\end{subfigure}
\begin{subfigure}{.5\textwidth}
\centering
\includegraphics[width=0.8\linewidth]{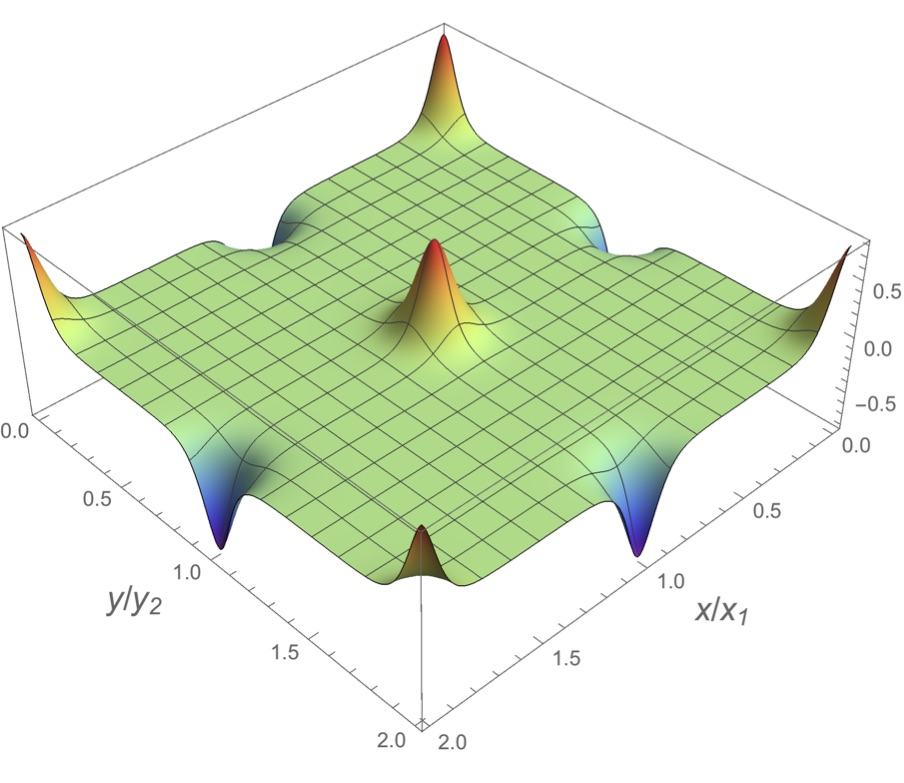}
\caption{$\chi$}
\end{subfigure}
\center
\begin{subfigure}{.5\textwidth}
\centering
\includegraphics[width=0.8\linewidth]{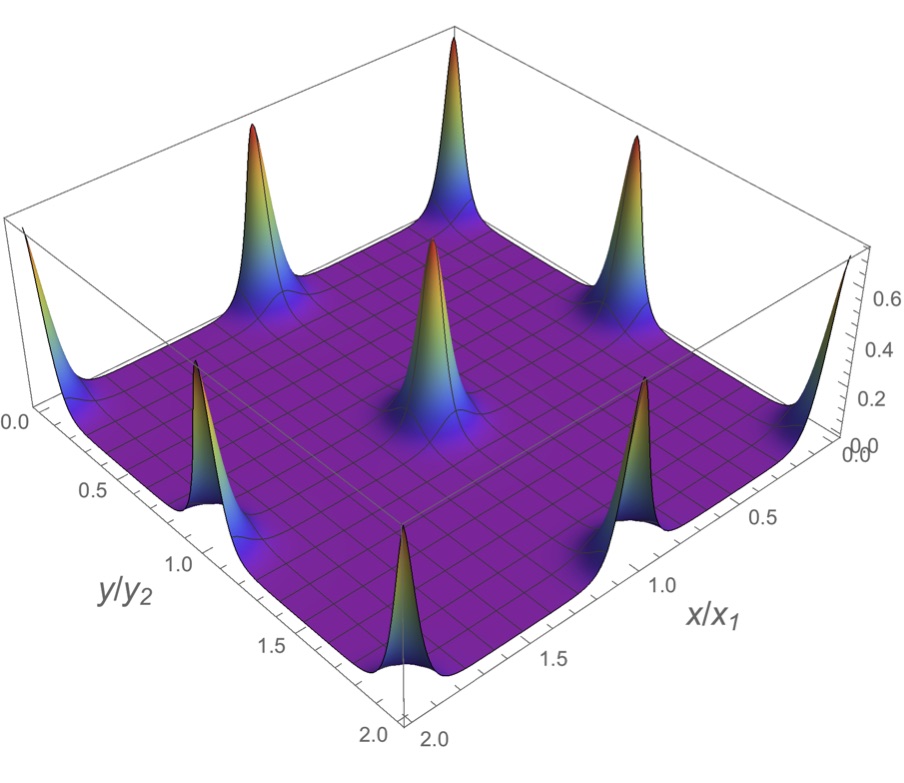}
\caption{$B$}
\end{subfigure}
\caption{Non-Abelian Square Vortex Lattice with anti-parallel $\chi$ field configurations at $x_1 = 20$ at $a =0.6$, $b=0.4$, $c=1.2$ and $\beta = 1.4 b/(c(c-1))$.}%
\label{figAntiParallel}%
\end{figure}

\begin{figure}[ptb]
\begin{subfigure}{.5\textwidth}
\centering
\includegraphics[width=0.8\linewidth]{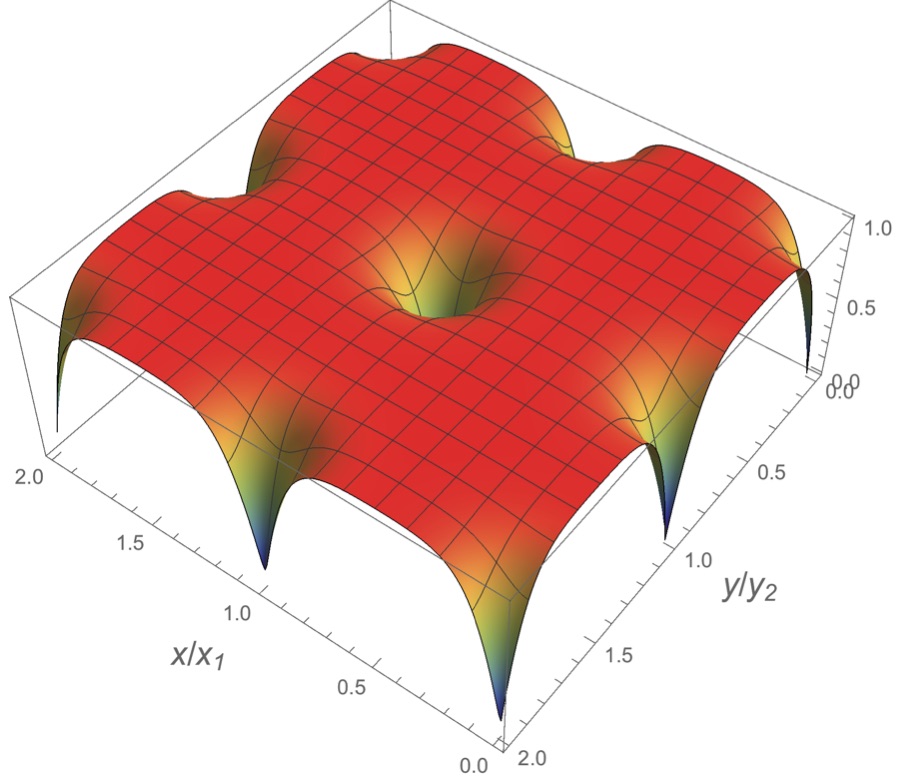}
\caption{$f$}
\end{subfigure}
\begin{subfigure}{.5\textwidth}
\centering
\includegraphics[width=0.8\linewidth]{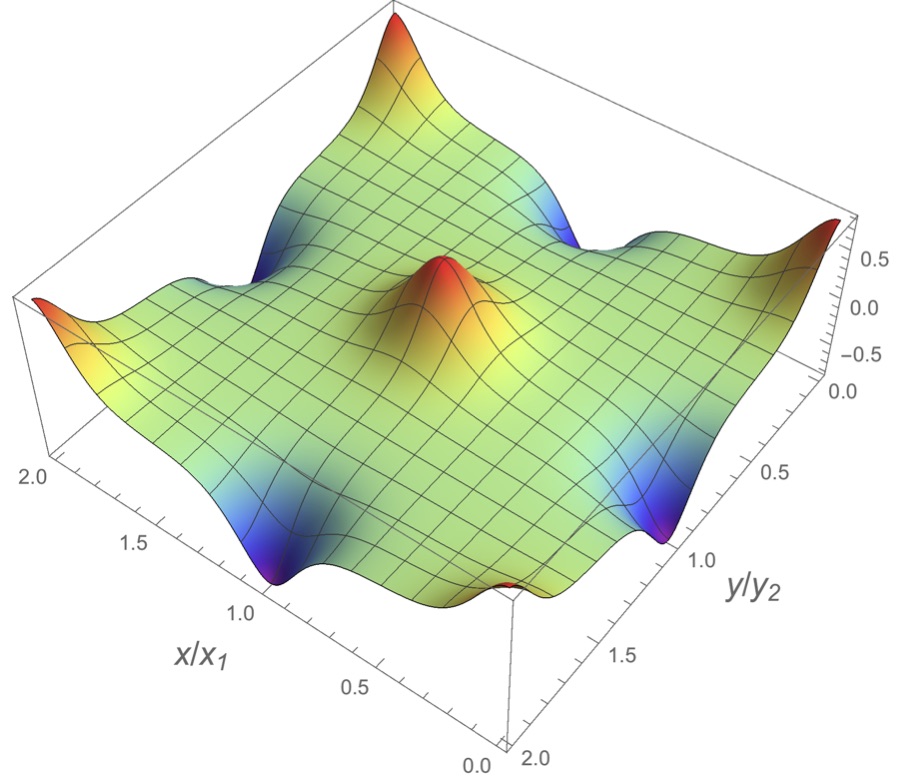}
\caption{$\chi$}
\end{subfigure}
\center
\begin{subfigure}{.5\textwidth}
\centering
\includegraphics[width=0.8\linewidth]{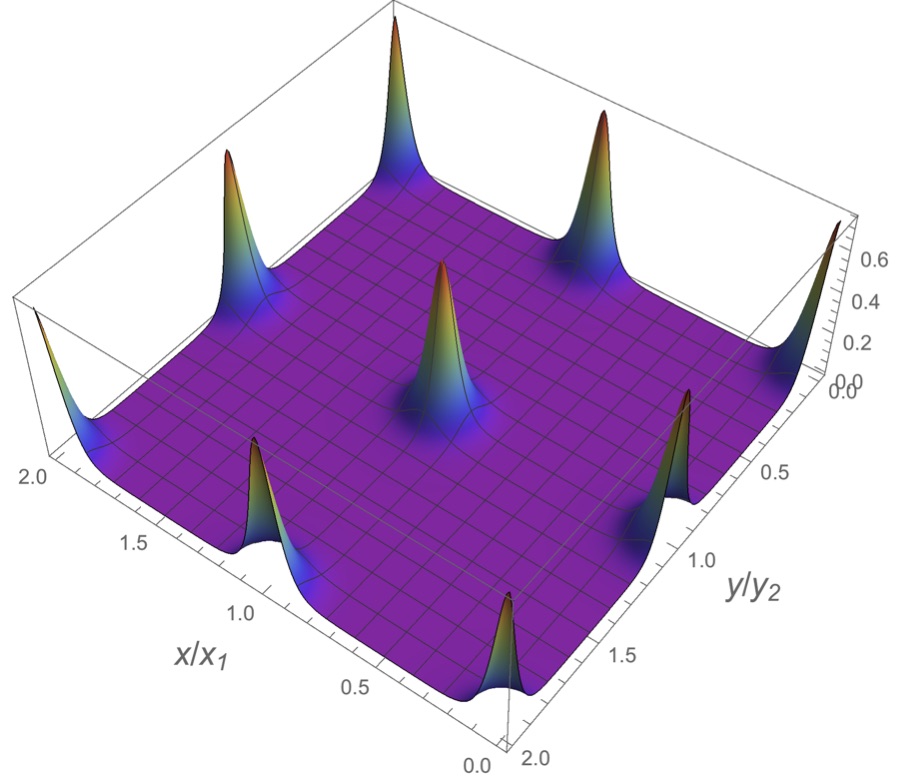}
\caption{$B$}
\end{subfigure}
\caption{Non-Abelian Square Vortex Lattice with anti-parallel $\chi$ field configurations at $x_1 = 10$ at $a =0.6$, $b=0.4$, $c=1.2$ and $\beta = 1.4 b/(c(c-1))$.}%
\label{figAntiParallel2}%
\end{figure}

In addition to the solutions we have found in the previous subsection, we have considered more general orientations of the $\chi$ field on the square lattice. Since we deal with a numerical relaxation procedure, we must make sure we start with an initial configuration which is close to an actual solution. To this end, it is difficult to imagine a case where the parallel configuration from lattice site to lattice site would not be the most stable state.  However, it may be possible that such alternative configurations could achieve some meta-stability for certain cases of the lattice spacing and constants.  For example, we are unable to rule out the possibility that the anti-parallel configuration of $\chi$ fields (analogous to an anti-ferromagnet) is metastable.  In figures (\ref{figAntiParallel}) and (\ref{figAntiParallel2}) we show the solutions for the $f$, $\chi$, and $B$ fields for two values of the lattice spacing.  These are generated using a similar ansatz for the $f$ and $B$ fields as the cases above, and an anti-parallel configuration for $\chi$.  

In order to test the stability of these anti-parallel configurations we may consider small perturbations of the $\vec{\chi}$ field, and observe the response.  For the case of solution shown in (\ref{figAntiParallel}) we find that small perturbations leave $\vec{\chi}$ disordered from site to site, with directors pointing in random directions.  This is as expected since the large spacing between lattice sites prevents interaction of the $\chi$ field localized at each site.  We conclude that in this particular case, no configuration of $\chi$ directors is preferred. This is simply a confirmation that, for large lattice spacing (or small vortex interaction) the system has independent orientational moduli on each vortex site (see the following section).

\begin{figure}[ptb]
\begin{subfigure}{.5\textwidth}
\centering
\includegraphics[width=0.7\linewidth]{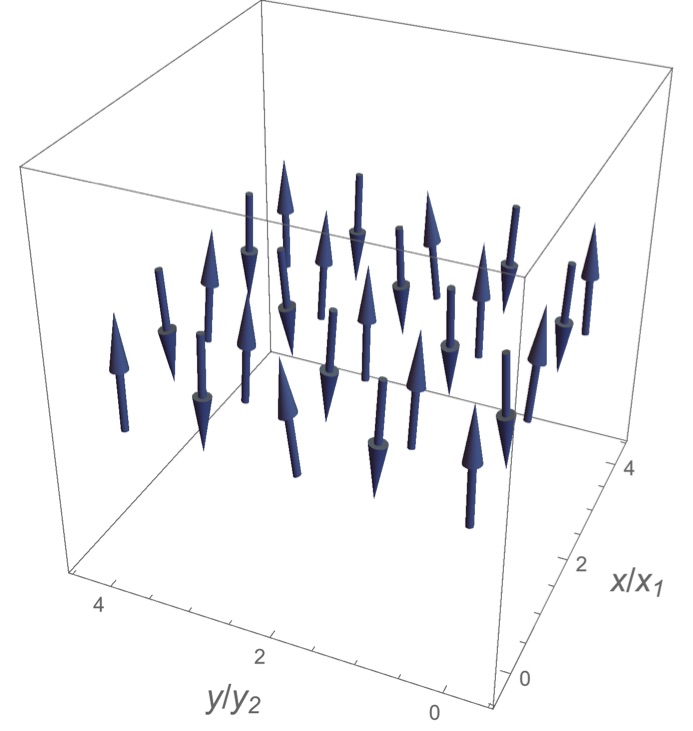}
\caption{1000 steps.}
\end{subfigure}
\begin{subfigure}{.5\textwidth}
\centering
\includegraphics[width=0.7\linewidth]{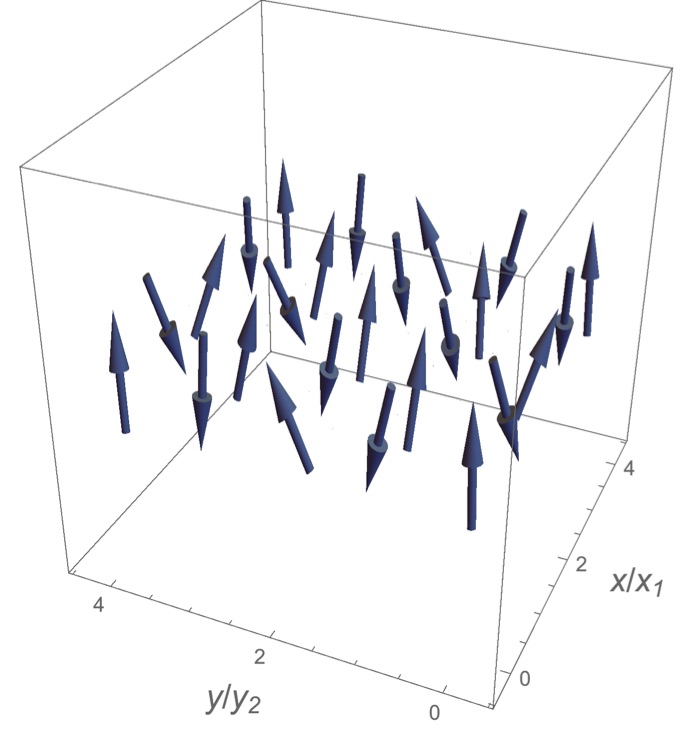}
\caption{2000 steps.}
\end{subfigure}
\begin{subfigure}{.5\textwidth}
\centering
\includegraphics[width=0.7\linewidth]{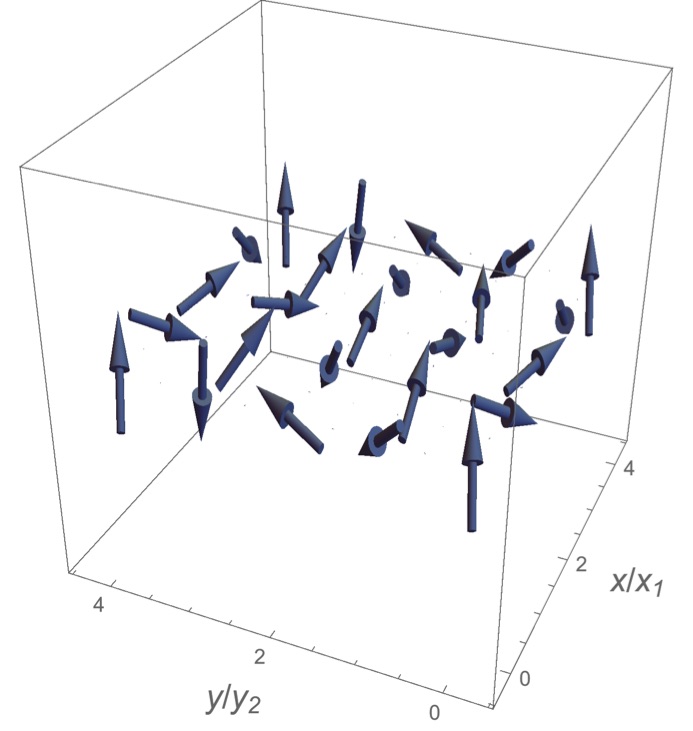}
\caption{3000 steps.}
\end{subfigure}
\begin{subfigure}{.5\textwidth}
\centering
\includegraphics[width=0.7\linewidth]{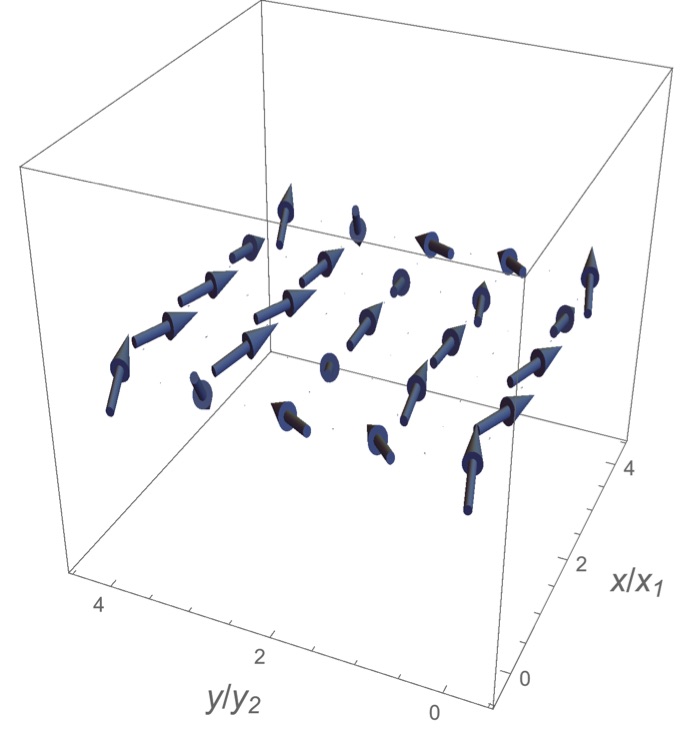}
\caption{4000 steps.}
\end{subfigure}
\begin{subfigure}{.5\textwidth}
\centering
\includegraphics[width=0.7\linewidth]{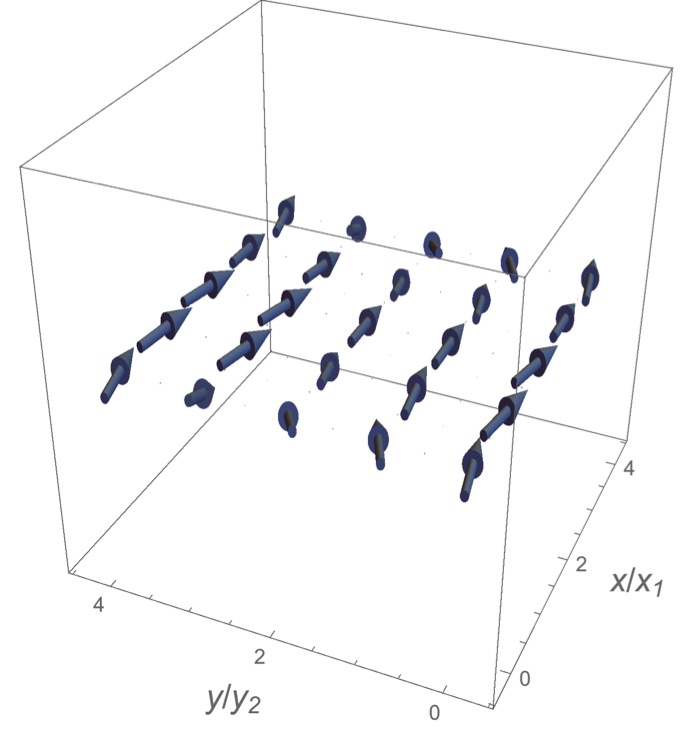}
\caption{5000 steps.}
\end{subfigure}
\begin{subfigure}{.5\textwidth}
\centering
\includegraphics[width=0.7\linewidth]{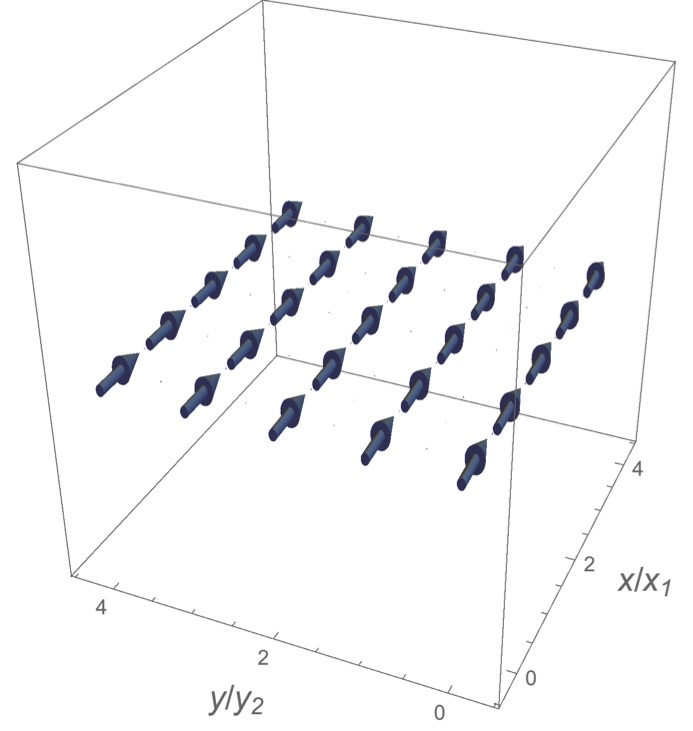}
\caption{Converged.}
\end{subfigure}
\caption{Time step evolution of the $\vec{\chi}$ field in the initially anti-parallel configuration at lattice sites with lattice spacing $x_1$ = 10.  The perturbation was given by introducing a small vector wave on the $\vec{\chi}$ field with Fourier coefficients generated randomly.}%
\label{figAntiParallelEvolution}%
\end{figure}

On the other hand, for closer lattice spacings, such as the case presented in (\ref{figAntiParallel2}), we find that a small perturbation of the initial anti-parallel configuration, leads to an instability.  In this case, the $\chi$ directors reorient in the parallel configuration.  In figure (\ref{figAntiParallelEvolution}) we illustrate the field configuration at various time steps in the minimization procedure.  We generate the initial perturbation of $\vec{\chi}$ by introducing a small wave form in $\vec{\chi}$ with Fourier coefficients generated randomly.  We conclude that the anti-parallel configuration is not stable in this case either.  

Of course we cannot argue on general grounds that stable anti-parallel configurations for $\vec{\chi}$ may exist for particular ranges of the parameter space and lattice spacings.  For all cases we have considered we find that the either the parallel or the disordered configuration for the $\vec{\chi}$ field are the only stable solutions, depending on lattice spacing.

\subsection{Searching for other meta-stable configurations}

\begin{figure}[ptb]
\begin{subfigure}{.5\textwidth}
\centering
\includegraphics[width=0.7\linewidth]{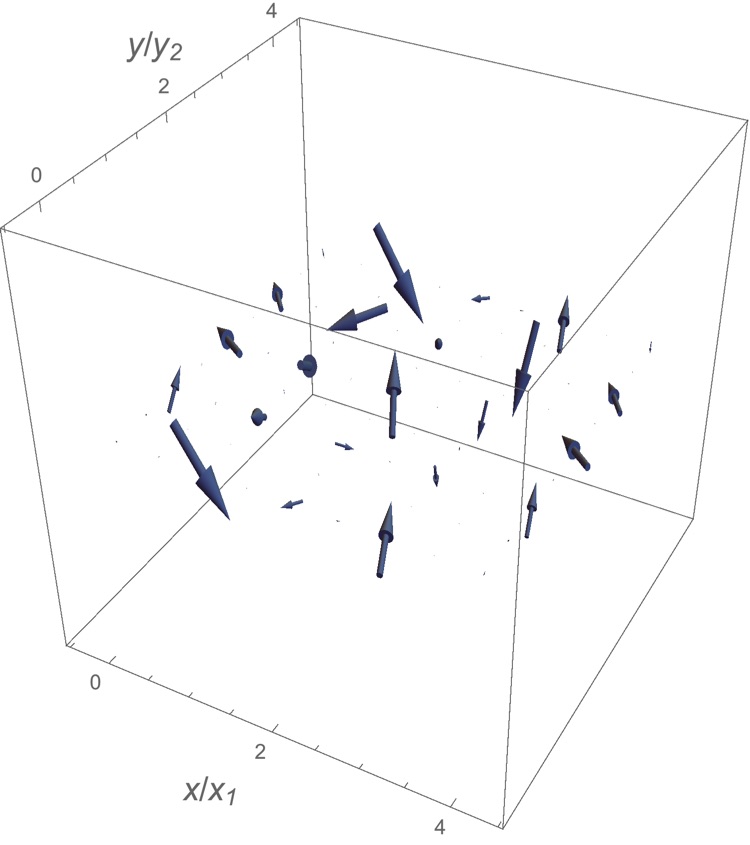}
\caption{0 steps.}
\end{subfigure}
\begin{subfigure}{.5\textwidth}
\centering
\includegraphics[width=0.7\linewidth]{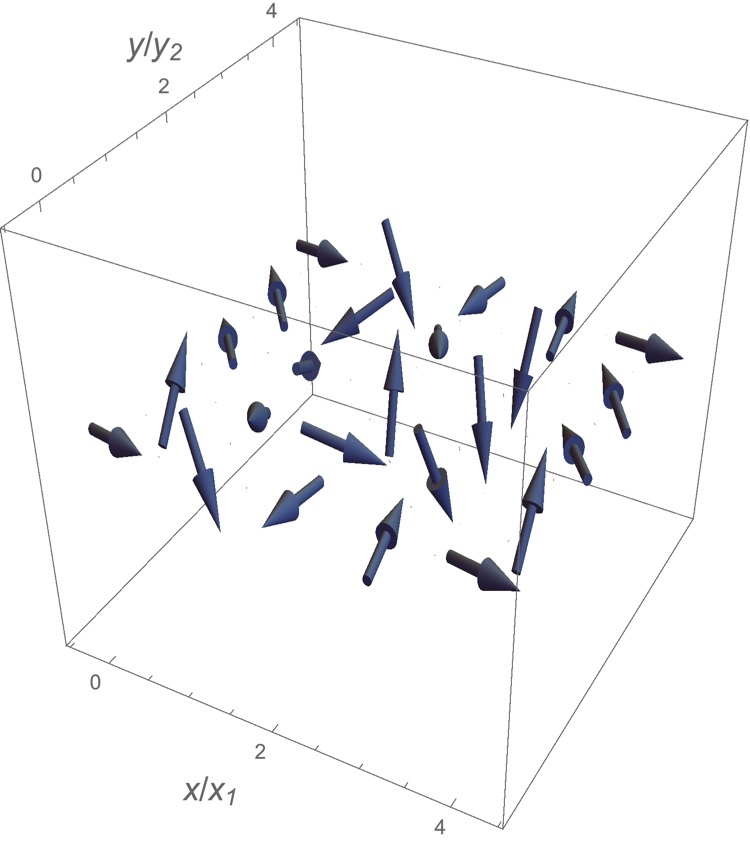}
\caption{500 steps.}
\end{subfigure}
\begin{subfigure}{.5\textwidth}
\centering
\includegraphics[width=0.7\linewidth]{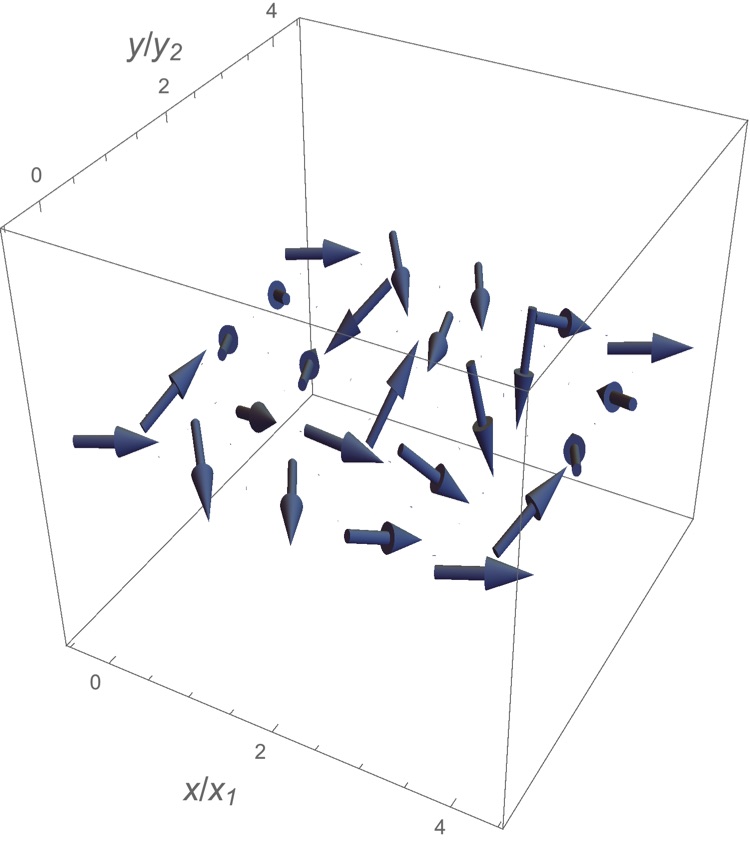}
\caption{2000 steps.}
\end{subfigure}
\begin{subfigure}{.5\textwidth}
\centering
\includegraphics[width=0.7\linewidth]{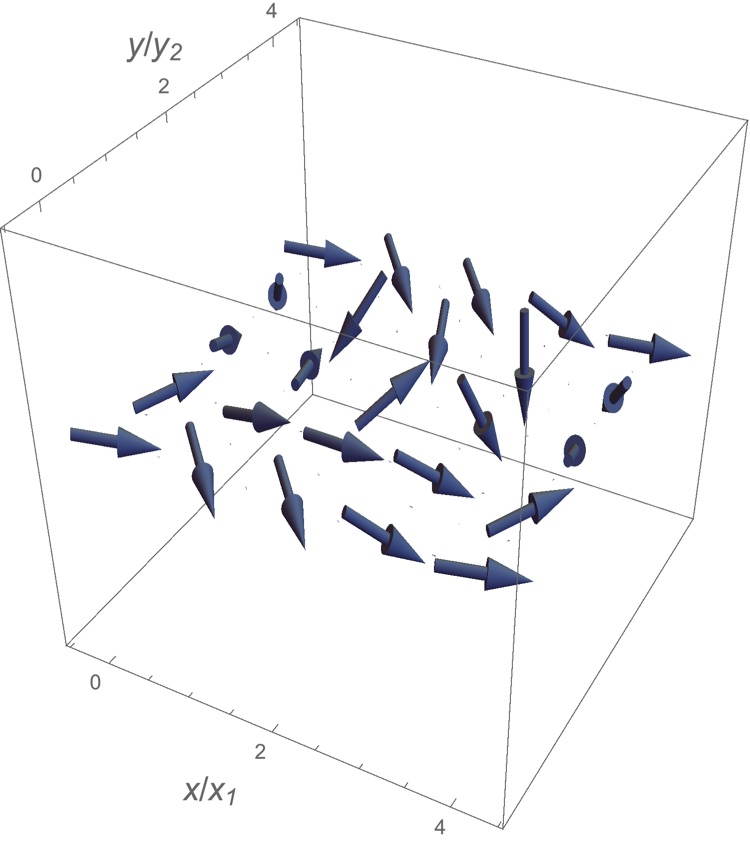}
\caption{3000 steps.}
\end{subfigure}
\begin{subfigure}{.5\textwidth}
\centering
\includegraphics[width=0.7\linewidth]{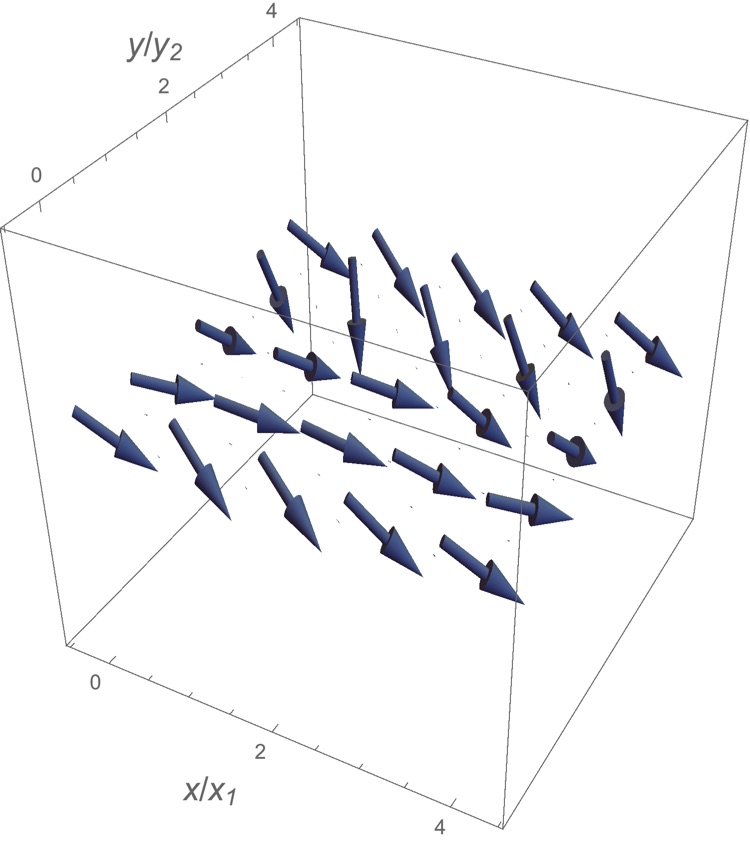}
\caption{5000 steps.}
\end{subfigure}
\begin{subfigure}{.5\textwidth}
\centering
\includegraphics[width=0.7\linewidth]{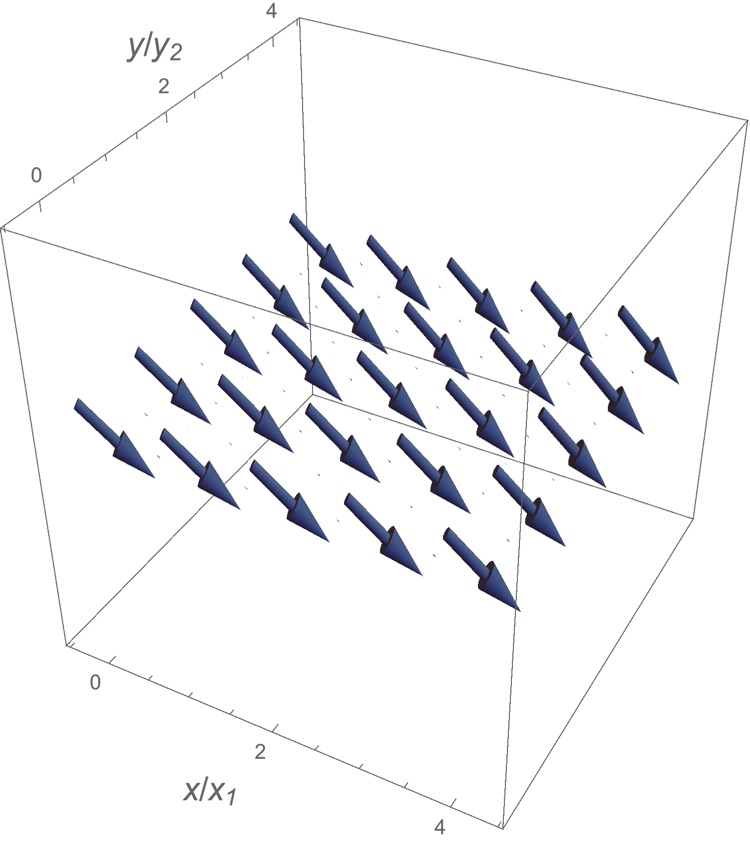}
\caption{Converged}
\end{subfigure}
\caption{An example of a randomly generated $\vec{\chi}$ field relaxing to the parallel configuration.  Here $x_1 = 10$.}%
\label{figRandEvolution}%
\end{figure}

In an effort to be more complete with the solution space for the $\vec{\chi}$ field configuration, we attempted to find additional configurations for the $\vec{\chi}$ lattice. A priori it is hard to imagine any relaxation seed besides the parallel or anti-parallel configurations, therefore we resorted to a more statistical approach based on random initial orientations per lattice site, repeating this many times for different configurations. This was done by Fourier decomposing the $\vec{\chi}$ field with randomly generated coefficients.  The minimization procedure was then carried out with this initial ansatz for $\vec{\chi}$, and was continued until convergence was achieved.  This procedure was repeated several times for several different values of the parameter space.  In figure (\ref{figRandEvolution}) we show the relaxation procedure for a particular initial $\vec{\chi}$ configuration.  In all cases considered, the $\vec{\chi}$ configuration relaxed to either a disordered state, or a parallel configuration depending on the spacing between lattice sites.  As previously mentioned, this is no surprise when interaction strength between lattice sites are considered.

We mention that this analysis is far from complete, and the possibility of parameter ranges where other metastable configurations appearing cannot be ruled out. In order to find them however, it is necessary to start with an intelligent seed and hence a clear idea of what these might look like.
\section{Energy}

In this section we present the numerical results regarding the energetics of our solutions. We compare the energies of square and triangular lattice configurations and, more importantly, those with and without the condensation of the $\chi$ field in or around the vortex cores. In the dimensionless units used, the energy functional is 
\be
E = E^f + E^\chi,
\ee
where
\be
E^f = \int d^2x \left[(\nabla f)^2 +B^2+\frac{1}{2}\left(f^2-1\right)^2+a f^2 \left(Q_x^2+Q_y^2\right)\right],
\ee
and
\be
E^\chi = \int d^2x \left[\frac{1}{2c\beta}(\nabla \chi)^2+\frac{1}{\beta}\frac{b}{c(c-1)}\left(\frac{1}{2}\chi^4-\chi^2(1-c f^2)\right)\right].
\ee

In the absence of the $\chi$ sector it is a well-known result that the triangular lattice has lower energy per unit cell compared to the square one. We have indeed confirmed this numerically. For example, for the solutions shown in figure (\ref{fig1}), comparing the energies per unit cell we obtained
\be
E^f_\square = 6.78, \quad E^f_\vartriangle = 6.57,
\ee

where the subscript on the energy indicates the geometry of the lattice. The energy of the triangular geometry is lower, as is well known. This result holds in general for all the ranges of parameters considered in this paper, we present some representative data in the table shown in figure \ref{figtable}. \newline

A more interesting and new comparison is that between Abrikosov lattices and non-Abelian vortex lattices. The energy for the solution shown in Figure \ref{fig2}, for example, is $E_2 = 7.991$. The corresponding energy without $\chi$ is $E_2^{\chi=0} = 7.999$. Similarly, the solution of figure \ref{fig3} has energy $E_3 = 7.988$ and that without $\chi$ (that of the corresponding pure Abrikosov lattice) $E_2^{\chi=0} = 7.997$. Therefore, once again, even in the presence of the additional field, the triangular lattice has least energy per unit cell. Note that it is energetically favourable for the lattice to nucleate the $\chi$ field in the vortex cores. This is not surprising, these solutions represent an ideal non-Abelian vortex lattice with well separated non-Abelian vortices. These vortices have lower energy than the corresponding ANO vortices \cite{Shifman:2014oqa}, a lattice of them should therefore also have lower energy if these don't interact. Interestingly the energy is lower but not significantly so, this was also found in \cite{Shifman:2014oqa} which allowed an approximate kink solution to be constructed. \newline

In this system the non-Abelian vortex lattices of triangular geometries are therefore the lowest energy solutions, in the limit of large separation of vortices. This result holds for all numerical values of the parameters we investigated, and seems to be a general result of this setup. We present some numerical data of representative solutions in the table shown in figure \ref{figtable2}.  \newline

\begin{figure}[ptb]
\centering
\begin{tabular}{|c|c|c|c|c|c|c|c|}
\hline
$a$&$b$&$\beta$&$x_1$&$E^\chi_{\square}$&$E^\chi_{\triangle}$&$E^{\chi=0}_{\square}$&$E^{\chi=0}_{\triangle}$\\ \hline
0.9&0.4&2.33&20&7.314&7.308&7.999&7.997\\\hline
0.9&0.7&4.08&20&7.11&7.10&8.000&7.997\\\hline
0.6&0.7&4.08&20&5.523&5.518&5.772&5.769\\\hline
\end{tabular}
\caption{Representative energies at different parameters and lattice spacings for square or triangular lattices of the non-Abelian type. $c=1.07$.}%
\label{figtable2}%
\end{figure}

When the non-Abelian vortices are more tightly packed, for smaller lattice spacing, we must verify whether the solutions with a delocalised $\chi$ field over the whole lattice are actually energetically preferred over solutions in which $\chi$ vanishes. We numerically checked that this was the case, some representative values are shown in the table in figure \ref{figtable3} (these lattice spacing values correspond to the solutions shown in figures \ref{fig4} and \ref{figtr}). Therefore, for all lattice spacings, even when the $\chi$ field does not vanish outside of the vortex cores, the system always prefers to have the orientational field than not.

\begin{figure}[ptb]
\centering
\begin{tabular}{|c|c|c|c|c|}
\hline
$x_1$&$E^\chi_{\square}$&$E^\chi_{\triangle}$&$E^{\chi=0}_{\square}$&$E^{\chi=0}_{\triangle}$\\ \hline
20&7.11&7.10&8.000&7.997\\\hline
10&6.89&6.83&7.81&7.76\\\hline
8&6.68&6.59&7.62&7.53\\\hline
\end{tabular}
\caption{Representative energies at different parameters and lattice spacings for square or triangular lattices of the non-Abelian type. $a=0.9$, $b=0.7$, $\beta= 4.08$, $c=1.07$.}%
\label{figtable3}%
\end{figure}

\subsection{Phase diagram of ideal non-Abelian vortex lattices}

As shown in section (\ref{int}) the condition of type II superconducting vortices changes in the presence of the $\chi$ field. In particular, we must check for what values of the parameters it is energetically convenient for a non-Abelian vortex of flux $\Phi$ to form a lattice of $n$ vortices of flux $\Phi/n$. Only then will the lattice formation be preferred for this type of superconductor. This section is devoted to carrying out such a numerical analysis and therefore mapping out the phase space of this system. This will determine the kind of superconductor that exists (type I or type II) as a function of parameters. \newline

Our strategy is the following: we solve the isolated vortex ODEs for non - Abelian vortices as a function of the parameters $a$ and $b$ for a single vortex of 2 flux quanta and calculate its energy, we then compare this energy to that of 2 vortices of a single flux quantum, at the same parameter values assuming no interaction (large separation). If the energy of the latter is smaller, then the system will want to be in a vortex lattice state and is a superconductor of type II. As discussed previously, the transition to this kind of superconductivity for the standard Abrikosov system (corresponding to $\chi=0$ or equivalently $b=0$), happens at $a=1$, when the gauge boson and scalar field masses are equal. This is commonly known as the BPS point, where the vortex interaction energy vanishes. For this comparison, we want to keep the flux constant and vary only the parameters of the equations. Therefore we will switch back to the rescaled convention adopted in section (\ref{int}). In this convention, the quantum flux is $\Phi = 2\pi \sqrt{2} n$, and importantly is independent of $a$ (this in turns makes the upper critical field also independent of $a$). \newline

To study isolated non-Abelian vortices we must go back to equations (\ref{chieqs1}) - (\ref{chieqs3}) and solve them with the boundary conditions
\be
f(0) = 0,\quad f(\infty)=1,
\ee
\be
\chi'(0) = 0,\quad \chi(\infty)=0,
\ee
\be
Q(0) = \sqrt{2} n,\quad Q(\infty)=0.
\ee

\begin{figure}[ptb]
\centering
\includegraphics[width=0.5\linewidth]{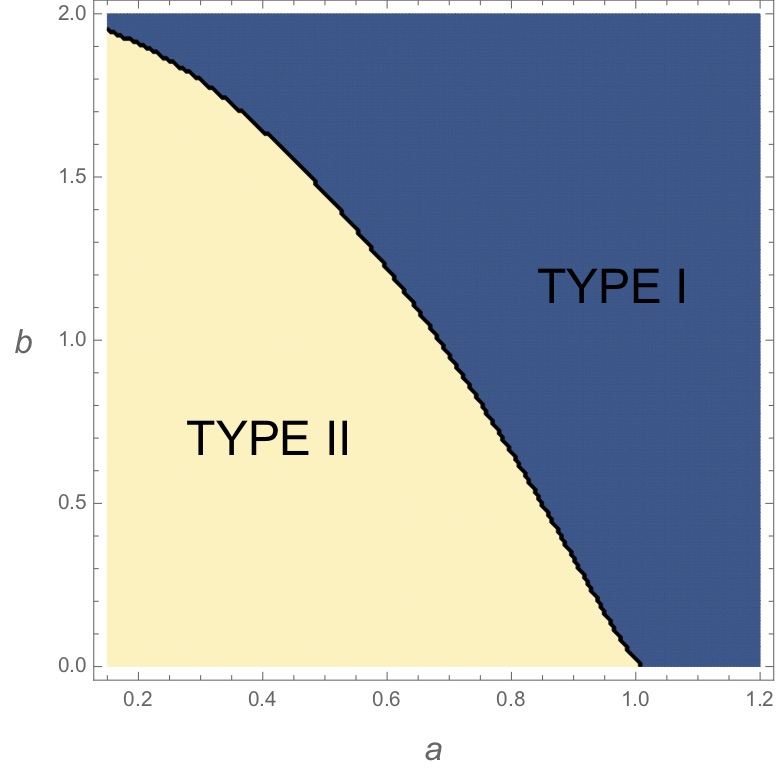}

\caption{Superconductivity type as a function of parameters $a$ and $b$ at fixed flux for the non-Abelian vortices at $c=1.07$ and $\beta = 26.8$.}%
\label{figx}%
\end{figure}

Our comparison is therefore between energies of solutions with $n=2$ and $n=1$ as we vary $a$ and $b$. The result of this analysis is shown in figure \ref{figx}. We can distinguish two main sections of this plot. The section marked II in this plot corresponds to a region in which the energy of two separated vortices of a single flux is less than that of a single vortex of flux 2. This region corresponds to type II superconductivity. We see immediately that for $b=0$, which corresponds to the usual Abrikosov vortex with $\chi=0$, the critical point of this phase is at $a=1$, as discussed in section (\ref{othereq}). As we vary $b$ we enter the non-Abelian vortex solutions. Here we find that there is a whole dome of parameter space of type II superconductive behaviour. We can label this as non-Abelian type II superconductivity. Beyond this region, in the region on the plot called I, the high flux single vortex solution is energetically preferred and the system is no longer of type II. It appears that as we increase $b$, and make the system increasingly more non-Abelian (in the sense of increasing the core value of $\chi$), the critical point of transition in the $a$ parameter is lowered, i.e. the extra scalar field means criticality is achieved at higher Higgs scalar masses, and no longer when this is equal to the gauge boson mass. This was the result anticipated in the discussion of section (\ref{othereq}), but it is now mapped numerically. The additional scalar channel provides an attractive force between the vortices, hence the $\chi$ field pushes the system towards type I.  Note that the upper value of $b \approx 2$ here is actually physical in the sense that above this line we violate the vacuum condition (\ref{cond2}) and would enter a phase in which the $\chi$ field wants to condense in the vacuum. We repeated this analysis for several values of $c$ and $\beta$ and find similar results for the shape of the critical line. The main difference being that the physical limit line on the $b$ axis changes as one changes $c$ or $\beta$. \newline

This diagram illustrates neatly where the solutions discussed in section 2 (which adopted the other convention) actually exist. They are representatives of the energy minimizing solutions, at fixed flux, that exist in the region of type II non-Abelian superconductivity. \newline

\begin{figure}[ptb]
\begin{subfigure}{.5\textwidth}
\centering
\includegraphics[width=0.8\linewidth]{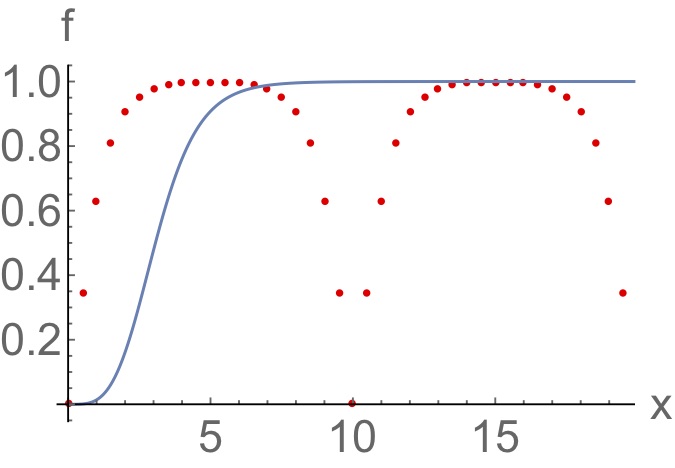}
\caption{$f$}
\end{subfigure}
\begin{subfigure}{.5\textwidth}
\centering
\includegraphics[width=0.8\linewidth]{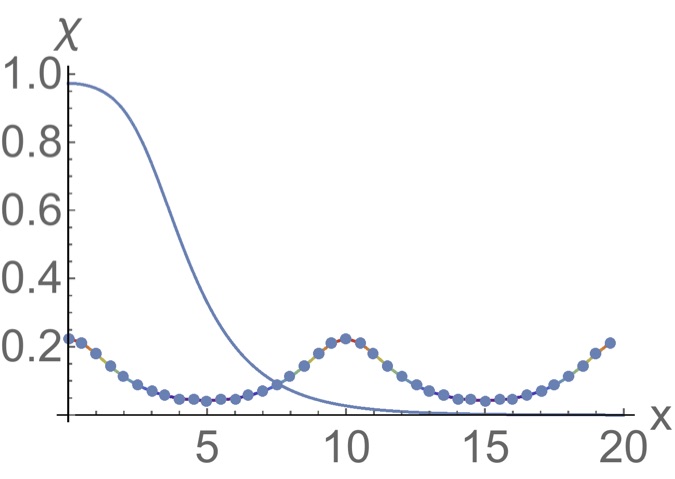}
\caption{$\chi$}
\end{subfigure}
\caption{Scalar field and $\chi$ field profiles comparing lattice (dotted line) and isolated vortex (solid line) of same total flux at $a=0.53$, $x_1 = 20$, $b=0.1$, $c=1.2$ and $\beta = 1.4 b/(c(c-1))$.}%
\label{fig8}%
\end{figure}

\begin{figure}[ptb]
\centering
\includegraphics[width=0.5\linewidth]{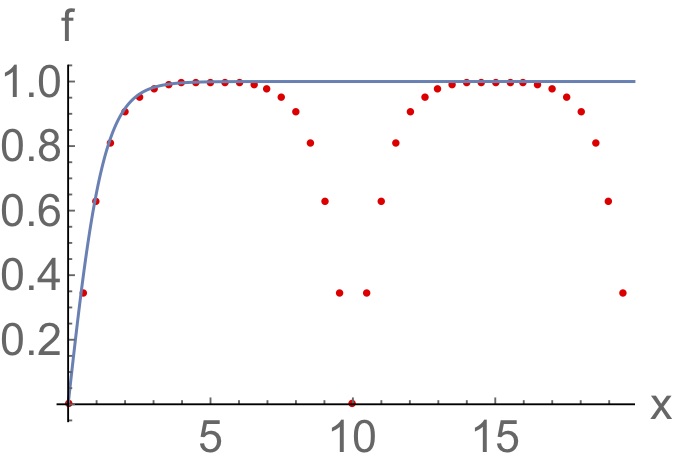}
\caption{The field profile for $f$ for an isolated vortex (solid line) is shown with the profile for the lattice (dotted).  This illustrates the consistency of the numerical procedure.}%
\label{fig6}%
\end{figure}

Finally, as an aside, we re-instanted the full 2D convention and also used the isolated vortex to compare our numerical lattice solutions with the 1D radial plots. First we compare the lattice solutions with single vortices carrying the total flux of the unit cell, this is shown in figure \ref{fig8}. In this figure the single vortex (solid line) carries 4 flux quanta, while the isolated vortices of the lattice each carry just one. In figure \ref{fig6} we show the isolated vortex (solid line) carrying the same flux as a vortex of the lattice (not the whole flux of the lattice). As seen by the figure there is a remarkable agreement between the isolated and lattice solutions demonstrating that our numerical procedure for the 2D solver reproduces the radial ODE solutions well.

\section{Low Energy theory}

The solutions represent lattices of non-Abelian vortices. In the core of the vortices, where $\psi=0$, the $\chi$ field condenses and breaks the $O(3)$ global symmetry down to $U(1)$, the rotations in the (1,2)-plane of internal space. If the lattice spacing is sufficiently large for the vortices to be well-separated, as per the solutions shown in figures $\ref{fig2}$ or $\ref{fig3}$, the low energy theory will be a lattice of $CP(1)$ non-linear sigma models localised on each vortex site.  In order to see this let us consider a small neighbourhood around a vortex core (defined approximately as a region of area in which $\chi$ is non-vanishing) in the $(x,y)$-plane and take the ansatz
\be
\chi^i =\sqrt{\frac{\mu^2}{2\beta}}\chi(x,y)S^i(t,z),
\ee
where $S^i(t,z)$ are orientational fields depending on the world sheet coordinates and satisfying $S^i S^i =1$. Inserting this ansatz into the action gives the low-energy action for the orientational moduli as
\be
S_o = \alpha \int dzdt \partial_k S^i \partial^k S^i, 
\ee
with $k = (t,z)$, and $\alpha = \frac{\mu^2}{2\beta} \int_v dxdy \chi^2$, where the integral runs over the neighbourhood of the vortex site. The overall action is then the sum of all the vortex site contributions so that
\be
S_{tot} =  \sum_i \alpha_i  \int dzdt \partial_k S_i^j \partial^k S_i^j,
\ee
where the index $i$ runs on over each vortex site and $\alpha_i$ refers to the integral $\alpha$ performed over connecting neighbourhoods of each vortex site. It is important to stress that for this kind of lattice each core moduli are independent to rotate on each vortex site since the $\chi$ fields are well localised. \newline

When the lattice spacing is lowered, such that the $\chi$ field delocalises from the vortex cores (see figure \ref{fig4}), only part of these orientational degrees of freedom survive as gapless excitations. These are the global rotations of the internal orientation throughout the whole lattice, namely locked rotations in internal space of all vortex sites. In this case each vortex site is not independent to rotate freely in internal space, the only gapless excitation is for each vortex to lock onto each other and rotate equally throughout the lattice. Then, the low energy theory is described by the $CP(1)$ model
\be
S = \int dzdt\; \partial_k S^j \partial^k S^j,
\ee 
where the integral now runs over the whole lattice.  This is similar to what happened for Skyrmions in \cite{Canfora:2016spb}.

\section{Conclusions}

This paper extends the previously known features of the Abrikosov string supporting non-Abelian moduli proposed in \cite{Shifman:2012vv}. In particular, it concludes the investigation on the nature of the superconducting regimes in it, with a complete map of superconducting type in parameter space. This model is a toy model of several physical systems, ranging from dark matter considerations \cite{Forgacs:2016dby} to confinement in QCD, wherever solitonic solutions of this type play an important role \cite{Eto:2006dx} \cite{Konishi:2001cj} \cite{Yung:2001cz}. Previous results in this model showed that such isolated vortices exist, and some studies on the properties of these vortices were made \cite{Shifman:2012vv} \cite{Shifman:2014oqa}. This paper extends the investigation to the full analysis of what type of superconductivity is involved, and what are the real low lying energy solutions the system adopts, in the case of parallel orientation. In particular, we first provide analytical evidence that the additional scalar field sector, providing the orientational degrees of freedom, acts as an additional attractive channel between isolated vortices. In turn, this modifies the usual $a=1$ BPS point of no vortex interaction energy. We then demonstrate numerically the existence of periodic arrays of parallely oriented non-Abelian vortices, both in the limit of large separations and when they are tightly packed, in square and triangular geometries. We map out the phase diagram of this superconductor and deduce the critical line of phase transition between type 2 and type 1 superconductivity in parameter space. When the system is in the type 2 region, the lowest energy solutions at fixed external flux and lattice spacing correspond to triangular non-Abelian vortex lattices. When the lattice spacing is large, the low energy theory of this system is described by an array of $CP(1)$ theories described by the local independent rotations of the orientational degrees of freedom at each lattice site. When the spacing is reduced, the self interactions of such degrees of freedom becomes important and the only remaining gapless excitation is a locked rotation of all lattice sites. This is similar to a ferromagnetic phase transition, in which independent spins couple into a single orientation under an applied magnetic field. With this in mind it would be interesting to allow all possible "spin" orientations of the vortices, rather than restricting to parallel ones. \newline 

We have considered alternate possible configurations of the spin lattice. For the ``anti-ferromagnetic" case (anti-aligned spins per vortex site) we determined that the configuration is unstable in the case of small lattice spacing. We also tried starting with random initial orientations and repeating the relaxation procedure with the hope that the system would spontaneously find alternative spin configurations. No such configurations were found. Based on this analysis we can conclude that the parallel orientation lattice is the most stable solution at small lattice spacings.  However, we admit that this analysis is far from concluded (alternative spin configurations might arise, for example, for lattice geometries besides the triangular or square case).  A more complete investigation of other potential stable configurations will have to wait for future projects.\newline

\subsection*{Acknowledgements}

G.T. would like to thank Tomas Andrade for preliminary work and the University of Oxford for kind hospitality. A.P thanks Universidad Adolfo Ibáñez for their kind hospitality. This work has been funded by a Fondecyt grant number 11160010. A.P. is funded by the NSERC Discovery Grant.

\begin{thebibliography}{99}      

\bibitem{Shifman:2012vv} 
  M.~Shifman,
  Phys.\ Rev.\ D {\bf 87}, no. 2, 025025 (2013)
  doi:10.1103/PhysRevD.87.025025
  [arXiv:1212.4823 [hep-th]].                                                                                         %

\bibitem{Auzzi:2003fs} 
  R.~Auzzi, S.~Bolognesi, J.~Evslin, K.~Konishi and A.~Yung,
  Nucl.\ Phys.\ B {\bf 673}, 187 (2003)
  doi:10.1016/j.nuclphysb.2003.09.029
  [hep-th/0307287].
  
\bibitem{Hanany:2003hp} 
  A.~Hanany and D.~Tong,
  JHEP {\bf 0307}, 037 (2003)
  doi:10.1088/1126-6708/2003/07/037
  [hep-th/0306150].
  
\bibitem{Hanany:2004ea}
  A.~Hanany and D.~Tong,
  JHEP {\bf 0404} (2004) 066
  doi:10.1088/1126-6708/2004/04/066
  [hep-th/0403158].
  
\bibitem{Shifman:2003uh} 
  M.~Shifman and A.~Yung,
  Phys.\ Rev.\ D {\bf 70}, 025013 (2004)
  doi:10.1103/PhysRevD.70.025013
  [hep-th/0312257].
\bibitem{Gorsky:2004ad}
  A.~Gorsky, M.~Shifman and A.~Yung,
  Phys.\ Rev.\ D {\bf 71} (2005) 045010
  doi:10.1103/PhysRevD.71.045010
  [hep-th/0412082].
  
\bibitem{Eto:2006pg} 
  M.~Eto, Y.~Isozumi, M.~Nitta, K.~Ohashi and N.~Sakai,
  J.\ Phys.\ A {\bf 39}, R315 (2006)
  doi:10.1088/0305-4470/39/26/R01
  [hep-th/0602170].
  
\bibitem{Shifman:2004dr} 
  M.~Shifman and A.~Yung,
  Phys.\ Rev.\ D {\bf 70}, 045004 (2004)
  doi:10.1103/PhysRevD.70.045004
  [hep-th/0403149].
  
\bibitem {shifman1}M. Shifman, "\textit{Advanced Topics in Quantum Field
Theory: A Lecture Course}" Cambridge University Press, (2012).

\bibitem {shifman2}M. Shifman, A. Yung, "\textit{Supersymmetric Solitons}"
Cambridge University Press, (2009).
\bibitem{Konishi:2002ky} 
  K.~Konishi,  10.1142/9789812776310,0024
  hep-th/0208222.
\bibitem{Konishi:2008vj} 
  K.~Konishi,
  Prog.\ Theor.\ Phys.\ Suppl.\  {\bf 177}, 83 (2009)
  doi:10.1143/PTPS.177.83
  [arXiv:0809.1370 [hep-th]].
\bibitem{Eto:2006dx} 
  M.~Eto, L.~Ferretti, K.~Konishi, G.~Marmorini, M.~Nitta, K.~Ohashi, W.~Vinci and N.~Yokoi,
  Nucl.\ Phys.\ B {\bf 780}, 161 (2007)
  doi:10.1016/j.nuclphysb.2007.03.040
  [hep-th/0611313].
\bibitem{'tHooft:1981ht} 
  G.~'t Hooft,
  Nucl.\ Phys.\ B {\bf 190}, 455 (1981).
  doi:10.1016/0550-3213(81)90442-9
\bibitem{Mandelstam:1974pi} 
  S.~Mandelstam,
  Phys.\ Rept.\  {\bf 23}, 245 (1976).
  doi:10.1016/0370-1573(76)90043-0
\bibitem{Witten:1984eb} 
  E.~Witten,
  Nucl.\ Phys.\ B {\bf 249}, 557 (1985).
  doi:10.1016/0550-3213(85)90022-7

  
\bibitem{Shifman:2013oia} 
  M.~Shifman and A.~Yung,
  Phys.\ Rev.\ Lett.\  {\bf 110}, no. 20, 201602 (2013)
  doi:10.1103/PhysRevLett.110.201602
  [arXiv:1303.7010 [hep-th]].
  
\bibitem{Monin:2013kza} 
  S.~Monin, M.~Shifman and A.~Yung,
  Phys.\ Rev.\ D {\bf 88}, no. 2, 025011 (2013)
  doi:10.1103/PhysRevD.88.025011
  [arXiv:1305.7292 [hep-th]].
  
\bibitem{Canfora:2016spb} 
  F.~Canfora and G.~Tallarita,
  Phys.\ Rev.\ D {\bf 94}, no. 2, 025037 (2016)
  doi:10.1103/PhysRevD.94.025037
  [arXiv:1607.04140 [hep-th]].

\bibitem{Peterson:2015tpa} 
  A.~J.~Peterson, M.~Shifman and G.~Tallarita,
  Annals Phys.\  {\bf 363}, 515 (2015)
  doi:10.1016/j.aop.2015.10.012
  [arXiv:1508.01490 [hep-th]].
\bibitem{Shifman:2015ama} 
  M.~Shifman, G.~Tallarita and A.~Yung,
  Phys.\ Rev.\ D {\bf 91}, no. 10, 105026 (2015)
  doi:10.1103/PhysRevD.91.105026
  [arXiv:1503.08684 [hep-th]].
\bibitem{Peterson:2014nma} 
  A.~Peterson, M.~Shifman and G.~Tallarita,
  Annals Phys.\  {\bf 353}, 48 (2014)
  doi:10.1016/j.aop.2014.11.001
  [arXiv:1409.1508 [hep-th]].
\bibitem{Shifman:2014oqa} 
  M.~Shifman, G.~Tallarita and A.~Yung,
  Int.\ J.\ Mod.\ Phys.\ A {\bf 29}, 1450062 (2014)
  doi:10.1142/S0217751X14500626
  [arXiv:1402.0733 [hep-th]].
  \bibitem{Tallarita:2015mca} 
  G.~Tallarita,
  Phys.\ Rev.\ D {\bf 93}, no. 6, 066011 (2016)
  doi:10.1103/PhysRevD.93.066011
  [arXiv:1510.06719 [hep-th]].
\bibitem{Forgacs:2016dby} 
  P.~Forgács and Á.~Lukács,
  Phys.\ Rev.\ D {\bf 95}, no. 3, 035003 (2017)
  doi:10.1103/PhysRevD.95.035003
  [arXiv:1612.03151 [hep-th]].
  \bibitem{Kobayashi:2013axa} 
  M.~Kobayashi, E.~Nakano and M.~Nitta,
  JHEP {\bf 1406}, 130 (2014)
  doi:10.1007/JHEP06(2014)130
  [arXiv:1311.2399 [hep-ph]].
  \bibitem{brandt1}
  E. H. Brandt, Phys. Rev. Lett. 78, 2208, 1997
  10.1103/PhysRevLett.78.2208
  \bibitem{brandt2}
  E. H. Brandt, Phys. Rev. B. 68,
  doi:10.1103/PhysRevB.68.054506
  [arXiv:0304237 [cond-mat]].
  \bibitem{brandt3}
  E. H. Brandt, Rep. Prog. Phys. 58, 1995 , 1465-1593.
  \bibitem{Manton:2004tk} 
  N.~S.~Manton and P.~Sutcliffe,``Topological solitons,''
  Cambridge University Press.
  \bibitem{Auzzi:2007wj} 
  R.~Auzzi, M.~Eto and W.~Vinci,
  JHEP {\bf 0802}, 100 (2008)
  doi:10.1088/1126-6708/2008/02/100
  [arXiv:0711.0116 [hep-th]].
  \bibitem{Bettencourt:1994kf}
  L.~M.~A.~Bettencourt and R.~J.~Rivers,
  Phys.\ Rev.\ D {\bf 51} (1995) 1842
  doi:10.1103/PhysRevD.51.1842
  [hep-ph/9405222].
  
\bibitem{Konishi:2001cj} 
  K.~Konishi and L.~Spanu,
  Int.\ J.\ Mod.\ Phys.\ A {\bf 18}, 249 (2003)
  doi:10.1142/S0217751X03011492
  [hep-th/0106175].
  
\bibitem{Yung:2001cz} 
  A.~Yung,
  In *Shifman, M. (ed.): At the frontier of particle physics, vol. 3* 1827-1857
\end{thebibliography}
\end{document}